\documentclass[twocolumn,prb,superscriptaddress,longbibliography]{revtex4-2}
\usepackage{graphicx}
\usepackage{color}
\usepackage{amsmath}
\usepackage{enumitem}
\usepackage{amssymb}
\usepackage{hyperref}
\usepackage[normalem]{ulem}
\usepackage{adjustbox}

\hypersetup{
	colorlinks = true,
	linkcolor = blue,
	citecolor = blue,
	urlcolor  = blue,
}

\newcommand{\be}{\begin{equation}}
\newcommand{\ee}{\end{equation}}

\newcommand{\bea}{\begin{eqnarray}}
\newcommand{\eea}{\end{eqnarray}}

\newcommand{\p}{\partial}

\newcommand{\tr}{{\rm \, tr\,}}

\renewcommand{\Im}{{\rm \, Im\,}}

\renewcommand{\vec}[1]{{\boldsymbol #1}}
\renewcommand{\epsilon}{\varepsilon}

\def\nn{\nonumber\\}

\renewcommand{\cite}[1]{[\onlinecite{#1}]}
\begin{document}
\title{Orbital susceptibility of T-graphene: Interplay of high-order van Hove singularities and Dirac cones}
\date{\today}

\author{D. O. Oriekhov}
\affiliation{Instituut-Lorentz, Universiteit Leiden, P.O. Box 9506, 2300 RA Leiden, The Netherlands}

\author{V. P. Gusynin}
\affiliation{Bogolyubov Institute for Theoretical Physics, National Academy of Science of Ukraine, 14-b Metrologichna Street, Kyiv, 03143, Ukraine}

\author{V. M. Loktev}
\affiliation{Bogolyubov Institute for Theoretical Physics, National Academy of Science of Ukraine, 14-b Metrologichna Street, Kyiv, 03143, Ukraine}
\affiliation{National Technical University of Ukraine KPI, 37 Peremogy Ave., Kyiv, 03056, Ukraine}

\begin{abstract}
 Square-octagon lattice underlies the description of a family of two-dimensional materials such as tetragraphene. In the present paper we show
 that the tight-binding model of square-octagon lattice contains both conventional and high-order van Hove points. In particular, the spectrum of
 the model contains flat lines along some directions composed of high-order saddle points. Their role is analyzed by calculating the orbital susceptibility
 of electrons. We find that the presence of van Hove singularities (VHS) of different kinds in the density of states leads to strong responses: paramagnetic for
 ordinary singularities and more complicated for high-order singularities. It is shown that at doping level of high-order VHS the orbital susceptibility 
 as a function of hoppings ratio $\alpha$ reveals the dia- to paramagnetic phase transition at $\alpha\approx0.94$. This is due to the competition of paramagnetic contribution of high-order VHS and diamagnetic contribution of Dirac cones.
 The results for the tight-binding model are compared with low-energy effective pseudospin-1 model near the three band touching point.
 \end{abstract}
\maketitle

\section{Introduction}
Possible existence of two new graphene allotropes, planar tetragraphene (or octagraphene) and buckled T-graphene composed of carbon octagons with tetrarings, was demonstrated some time ago using the Density Functional Theory (DFT) \cite{Liu2012PRL}. Several previous attempts to find such allotropes were made in Refs.\cite{Terrones2000,Enyashin2011}. It was noted that planar T-graphene allotrope should be the most stable one after graphene while the buckled T-graphene is not stable, and its fully relaxed state is very similar to planar T-graphene \cite{Kim2013}. Recently, the tetragraphene allotrope has been predicted to possess superconductivity with critical temperature up to around $20.8$ K \cite{QinyanGu:97401}.

Some geometrical and electronic properties, as well as low-energy physics of octagraphene were studied in Ref.\cite{Sheng2012}, the phase diagrams were analyzed and the existence of Mott metal-insulator phase transitions in the Hubbard model on square-octagon lattice was pointed out in \cite{Bao2014Nature,Yamashita2013PRB,Li2014,Pomata2019,Sun2015}.
In addition, structural and electronic properties of T-graphene and its modifications were studied by DFT calculations in Refs.\cite{Umemoto2010, Majidi2015,Majidi2017,Yin2013} and the kinetic stability with time was analyzed in Ref.\cite{Podlivaev2013}. Later, it was shown \cite{Gaikwad2017} that the 2D monolayers of Zn$_{2}$O$_{2}$ and Zn$_{4}$O$_{4}$ also have nearly ideal square-octagon lattice. In recent paper \cite{Gaikwad2020} the stability of multilayer materials such as ZnO composed of square-octagon lattice was studied
with the help of DFT technique. Also it was shown that MoS$_{2}$  transition metal dichalcogenide with square-octagon lattice can possess Dirac fermions with
Fermi velocity comparable to that of graphene \cite{Li2014}. The coexistence of Dirac fermions and nearly flat bands seems to be a very interesting
 property of square-octagon lattice and motivates us to study
physical quantities such as orbital susceptibility in terms on newly introduced concept of high-order van Hove singularities \cite{Yuan2019Nature}.

As is known, when the doping level approaches VHS, system can exhibit strong responses such as orbital paramagnetism in two-dimensional case \cite{Vignale1991PRL} or  chiral superconductivity in the case of graphene \cite{Nandkishore2012}. An {\it ordinary} VHS in two-dimensional electron system corresponds to logarithmic divergence of the density of states (DOS). The distinctive feature of high-order VHS is a more singular, power-law divergence of DOS with an asymmetric peak \cite{Yuan2019Nature,Isobe2019}. At the same time, the recent studies of two-dimensional lattices uncovered a wide family of exotic band structures \cite{Bradlyn2016} with flat bands and multi-band touching points, at which the quasiparticles are effectively described by high-pseudospin Hamiltonians.
 Flat bands can be considered as a limiting case of VHS with delta-function divergence of DOS.

 The prominent examples of materials with high-order VHS of different kind are bilayer graphene with tuned dispersion with the help of an interlayer voltage bias \cite{Shtyk2017}, $\mathrm{Sr}_{3} \mathrm{Ru}_{2} \mathrm{O}_{7}$ \cite{Efremov2019} and $\beta-\text{YbAlB}_{4}$ \cite{Ramires2012PRL}.
Recently it was also shown that when a high-order VHS is placed close to the Fermi level, density wave, Pomeranchuk orders, and superconductivity can all be enhanced \cite{Classen2020}. The role of high-order VHS on different types of instabilities in twisted bilayer graphene was analyzed in Ref.\cite{Lin2020arxiv}. The presence of van Hove singularities in twisted bilayer graphene \cite{Sherkunov2018} can lead to valley magnetism \cite{Chichinadze2020}, density waves and unconventional superconductivity \cite{Isobe2018PRX} such as topological and nematic superconductivity \cite{Wang2020arxiv}, the so-called "high-T$_{c}$" phase diagram \cite{Lin2019highTc}, Kohn-Luttinger superconductivity \cite{Gonzalez2019}.

The orbital susceptibility \cite{Ashcroft-book} measures the response of a time-reversal invariant electronic system to an external magnetic field.
To evaluate susceptibility of T-graphene analytically and numerically we use the formulas for susceptibility derived in Refs.\cite{Gomez-Santos2011PRL} and \cite{Raoux2015PRB}. We analyze the role of VHS of both kinds in orbital susceptibility for electrons on square-octagon lattice. Particularly, we show that
the flat lines in tight-binding band structure, which were firstly mentioned in Ref.\cite{Yamashita2013PRB}, also represent high-order VHS with inverse square
root divergence of DOS.

The paper is organized as follows. In Sec.\ref{sec:model} we describe the tight-binding Hamiltonian of square-octagon lattice. Then, in Sec.\ref{sec:spectrum-special} we derive effective low-energy Hamiltonians that describe bands around highly-symmetric points in Brillouin zone (BZ). Also we identify the type of VHS which are present in T-graphene. In Sec.\ref{sec:susceptibility} we perform numerical evaluation of susceptibility, and then analyze the qualitative physical effects of Dirac cones (Sec.\ref{sec:susceptibility-analytic-Yamashita}) and VHS using effective low-energy expansion (Secs. \ref{sec:4C} and \ref{sec:susceptibility-van-Hove}). The role of high-order VHS is discussed also in the Conclusions (Sec.\ref{Conclusion}) where we summarize the obtained results. In Appendix \ref{appendix:flat-lines} we analyze flat lines in the dispersion of middle bands, and in  Appendix \ref{sec:green-func} we present expressions for the Green's functions of tight-binding and
L\"{o}wdin Hamiltonians.

\section{Tight-binding model}
\label{sec:model}
\begin{figure}
	\centering
	\includegraphics[scale=0.75]{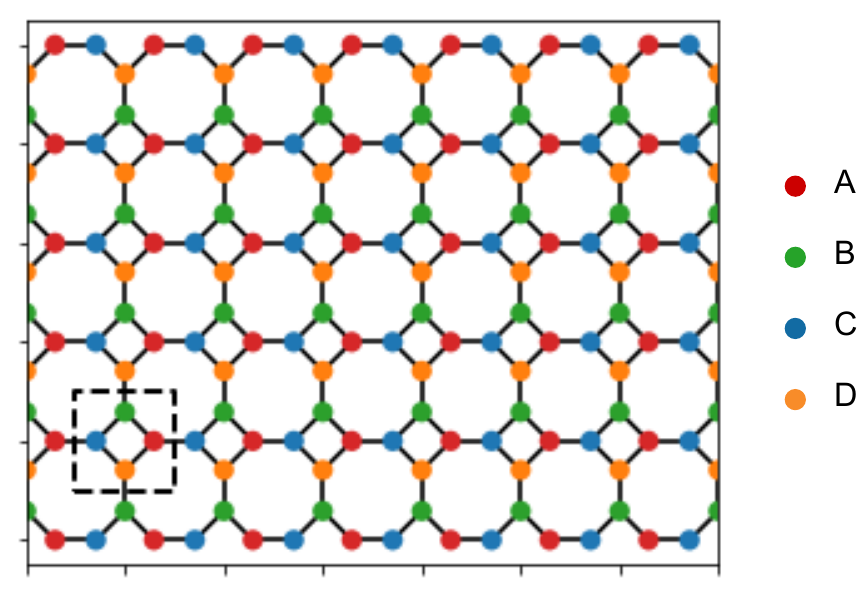}
	\caption{T-graphene lattice structure, which is described in main text. Each sublattice is denoted by its own color. Black dashed rectangle encircles one elementary cell. The hopping parameters between two small squares are $t_1$ and inside each small square - $t_2$.}
	\label{fig:Tgraphene-lattice}
\end{figure}
The square-octagon lattice consists of four atoms per unit cell which form a small square, and is shown on Fig.\ref{fig:Tgraphene-lattice}.  According to
Ref.\cite{Enyashin2011}, the numerical values  for all nearest neighbor interatomic distances are approximately equal to $1.429 \text{\AA}$ and lattice constant $a = 3.47 \text{\AA}$ for T-graphene. Ref.\cite{Sheng2012} gives the intra-square,  $1.48 \text{\AA}$, and inter-squares, $1.35 \text{\AA}$, distances, and similar values were reported in Ref.\cite{Gaikwad2020}. The basis vectors of Bravais lattice and reciprocal lattice are
\begin{align}
&\vec{a}_1=(a,\,\,0),\quad 	\vec{a}_2=(0, \,\,a);\nn
&\vec{b}_1=\left(0,\,\,\frac{2\pi}{a}\right),\quad 	\vec{b}_2=\left(\frac{2\pi}{a}, \,\,0\right).
\end{align}
In the tight-binding model, we take hopping parameters between atoms in two neighboring small squares to be $t_1$, and inside small square  $t_2$. The corresponding tight-binding Hamiltonian has the form \cite{Sheng2012,Yamashita2013PRB}
\begin{align}\label{eq:tight-binding-Hamiltonian}
H_{Tg}(\vec{k})=-\begin{pmatrix}
0 & t_2 & t_1e^{i k_x a} & t_2\\
t_2 & 0 & t_2 & t_1e^{i k_y a}\\
t_1e^{-i k_x a} & t_2 & 0 & t_2\\
t_2 & t_1e^{-i k_y a} & t_2 & 0
\end{pmatrix}.
\end{align}
and acts on the four-component wave functions $\psi=(\psi_A,\,\psi_B,\,\psi_C,\,\psi_D)$ (see Fig.\ref{fig:Tgraphene-lattice} for sublattice labels).
The above mentioned difference in interatomic distances can effectively described by tuning the hopping parameters $t_{1}$ and $t_{2}$. The values of these hopping parameters can be taken from DFT calculations: $t_1=2.9\,\text{eV}$ and $t_2=2.5\,\text{eV}$ were used in Ref.\cite{Sheng2012}, while $t_1=2.98\,\text{eV}$ and $t_2=2.68\,\text{eV}$ were found from DFT calculations inside one layer of octagraphene \cite{Li2020tg}.

\begin{figure*}
	\centering
	\includegraphics[scale=0.45]{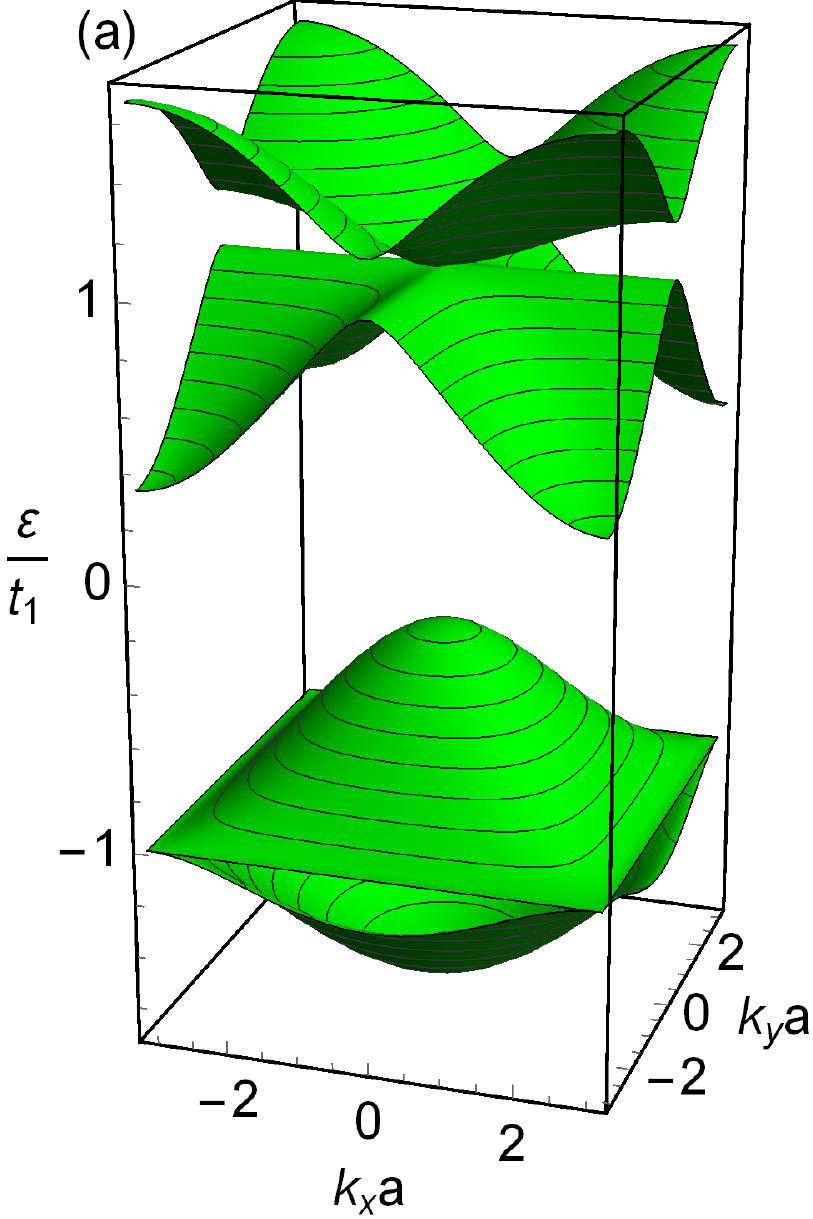}\qquad\qquad
	\includegraphics[scale=0.45]{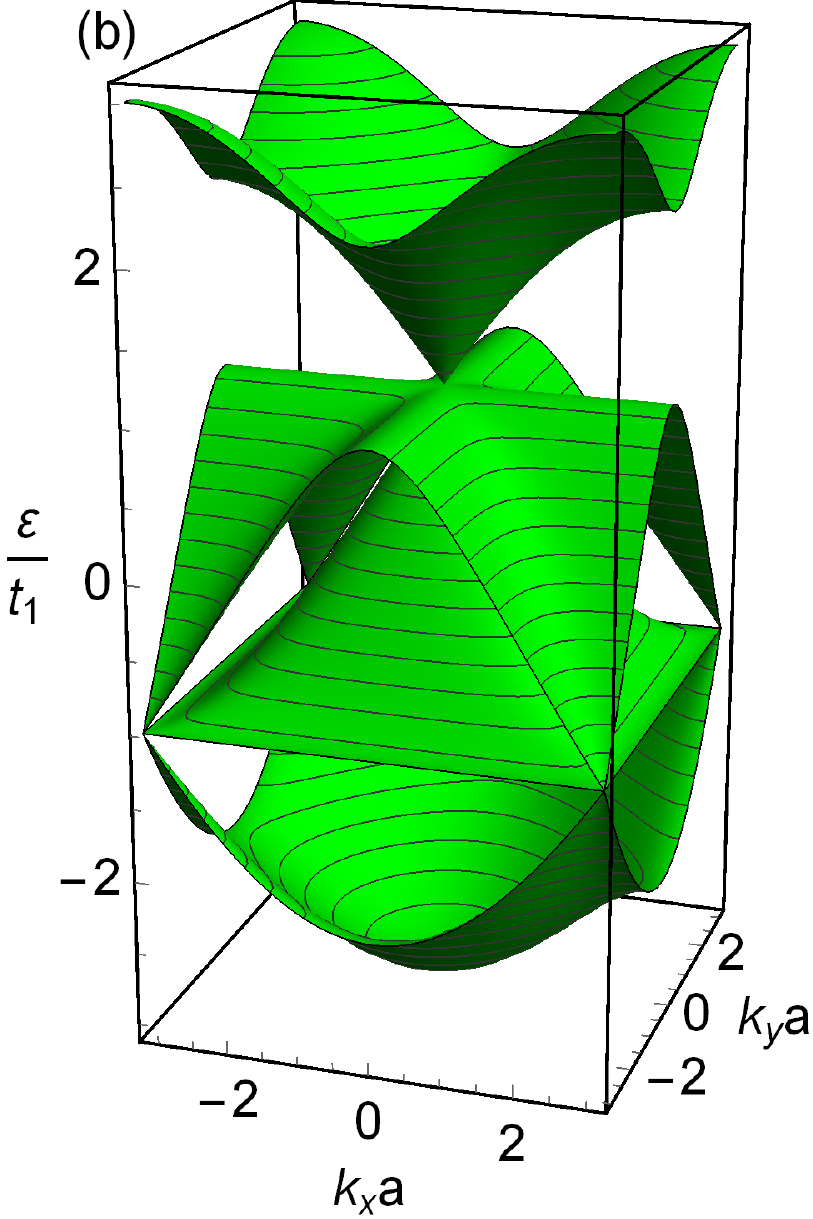}\qquad\qquad
	\includegraphics[scale=0.45]{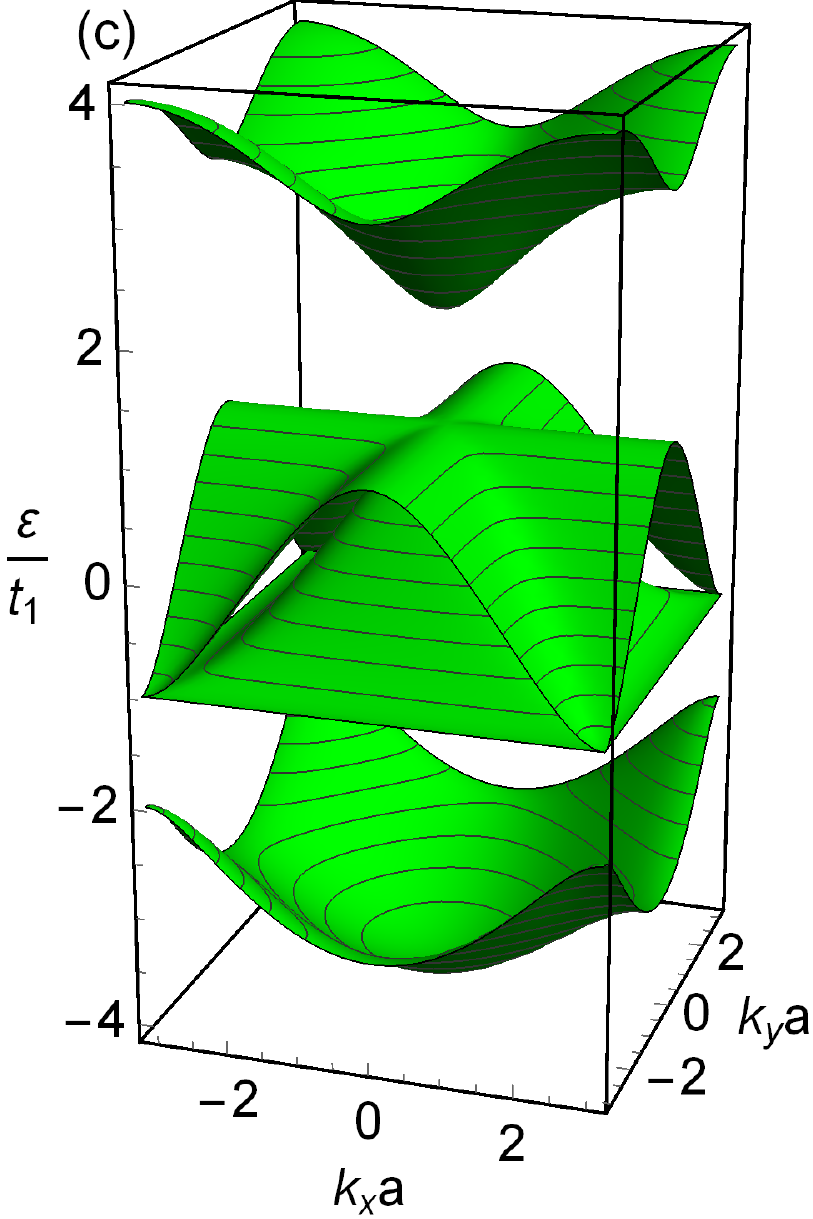}
	\caption{Spectrum which is given by Eq.\eqref{eq:spectral_eq} for three values of parameter $\alpha=t_2/t_1$: panel (a) $\alpha=\frac{1}{3}$, panel (b)
$\alpha=1$ and panel (c) $\alpha=\frac{3}{2}$. The energy $\epsilon$ is measured in units of hopping parameter $t_1$. On the panel (b) one can observe the three-band-touching points where the two Dirac cones meet nearly flat middle band. Black lines denote the lines of constant energies.}
	\label{fig:spectrum-moved}
\end{figure*}

The spectrum can be found from the equation $\det[\epsilon\mathbb{I}-H_{Tg}(\vec{k})]=0$, which after simplification reduces to \cite{Sheng2012,Yamashita2013PRB}
\begin{align}\label{eq:spectral_eq}
&\epsilon^4-2 \left(t_1^2+2 t_2^2\right) \epsilon^2+4 t_1 t_2^2 \epsilon  \left(\cos\left(a k_x\right)+\cos\left(a k_y\right)\right)-\nn
&-4 t_1^2 t_2^2 \cos\left(a k_x\right)\cos\left(a k_y\right)+t_1^4=0,
\end{align}
and has the form of depressed quartic equation. The spectrum is symmetric with respect to rotations on the angle $\frac{\pi}{4}$ in $k$-space, because the lattice
has a $C_4$ point symmetry group. Also the spectrum is symmetric with respect to transformations $\epsilon\to -\epsilon$
together with $k_{x}\to k_{x}\pm \frac{\pi}{a},\,k_{y}\to k_{y}\pm \frac{\pi}{a}$ (called chiral symmetry in \cite{Yamashita2013PRB}).
The Brillouin zone of square-octagon lattice
is a square with $-\frac{\pi}{a}<k_{x},\,k_y<\frac{\pi}{a}$. The corresponding highly-symmetric points are defined as \begin{align}\label{eq:GXM-points-def}
&\Gamma=(0,0),\quad M=\left(\pm\frac{\pi}{a},\pm\frac{\pi}{a}\right),\nn &X=\left(\pm\frac{\pi}{a},0\right),\,\,\left(0,\pm\frac{\pi}{a}\right),
\end{align}
and are located in the center, corners and the middle of each square site, respectively.
It is convenient to measure the energy in terms of $t_1$ hopping parameter, and introduce the dimensionless ratio of hopping parameters $\alpha=t_2/t_1$.
The 3D plots of the spectrum defined by Eq.\eqref{eq:spectral_eq} for several values of $\alpha$ are shown in Fig.\ref{fig:spectrum-moved}, while the 2D plots along highly-symmetric lines are represented in Fig.\ref{fig:spectrum-dos}. For $\alpha=1$, near the three-band-touching points $\Gamma$ and $M$, one observes almost flat middle bands \cite{Yamashita2013PRB}. These two middle bands support completely flat energy lines, which are extended over full BZ. Below we proceed with description of highly-symmetric points in terms of van Hove singularities in the DOS.

\section{Spectrum structure around highly-symmetric points: van Hove singularities}
\label{sec:spectrum-special}
Firstly, let us present general definitions that will be used throughout the text. By definition, the one-electron DOS per spin
is given by
\begin{align}\label{eq:dos-definition-int}
	D(\varepsilon)=\sum_{i=1}^{4} \int_{BZ} \frac{d^{2} k}{(2 \pi)^{2}} \delta\left[\varepsilon-\epsilon_{i}(\mathbf{k})\right],
\end{align}
with $i$ running over the band dispersions $\epsilon_{i}(\vec{k})$ found from Eq.\eqref{eq:spectral_eq}. Due to chiral symmetry the DOS is an even function
of energy. The ordinary VHS with the logarithmic diverging DOS occurs at saddle point $\vec{k}_{s}$ of a particular band in which
\begin{align}\label{eq:van-hove-ordinary}
\nabla_{\mathbf{k}}\epsilon(\mathbf{k})=\mathbf{0} \text { and } \operatorname{det} \mathcal{D}<0,
\end{align}
where $\mathcal{D}_{i j} \equiv \frac{1}{2} \partial_{i} \partial_{j} \epsilon(\mathbf{k})$ is the $2\times 2$ Hessian matrix of a dispersion $\epsilon(\mathbf{k})$ at $\vec{k}_s$. Here and below we use short-hand notation $\p_{i}=\p_{k_{i}}$. After proper rotation of a basis, the dispersion around saddle point can be conveniently represented as $\epsilon-\epsilon_{s}\approx-\zeta p_x^2 + \beta p_y^2$ with wave vector deviation  $\mathbf{p}=\mathbf{k}-\mathbf{k}_{s}$. The two coefficients $\zeta$ and $\beta$ are the eigenvalues of $\mathcal{D}$ and satisfy the condition $-\zeta \beta =\det \mathcal{D}<0$.

The high-order VHS corresponds to saddle point with the following properties \cite{Yuan2019Nature}:
\begin{align}
\nabla_{\mathbf{k}} \epsilon=\mathbf{0} \text { and } \operatorname{det} \mathcal{D}=0.
\end{align}
This class of VHS can be divided into two types: $\zeta=\beta=0$ (multicritical VHS), or $\zeta\neq 0,\,\,\beta=0$. The DOS is expected to have a power-law divergence
at such points. The position of all VHS can be found by differentiating Eq.(\ref{eq:spectral_eq}) and setting $\nabla_{\mathbf{k}} \epsilon=0$, from which we get
the system of equations:
\bea
&&\sin(ak_x)\left(\epsilon-t_1\cos(ak_y)\right)=0,\nonumber\\
&&\sin(ak_y)\left(\epsilon-t_1\cos(ak_x)\right)=0.
\eea

Below we perform expansion of the energy spectrum of T-graphene around highly-symmetric points and flat lines and identify the corresponding VHS type
with the DOS divergence.



\subsection{$\Gamma$ and $M$ points}
Before proceeding with calculation, we underline that previously mentioned symmetry of spectrum makes these two points equivalent up to change of energy sign.
Thus, the analysis around the $\Gamma$ point can be directly translated to the $M$ point and vice versa by chiral symmetry.

To find the approximate expressions for band energies around highly-symmetric points, we perform the series expansion of spectral equation \eqref{eq:spectral_eq}.
We write  $\epsilon=\epsilon_i^{(0)}+\delta$, with $\epsilon_i^{(0)}$ is the energy of i-th band exactly at the given point in k-space. Then, we expand equation
into series in $\delta$ and $\vec{k}a$ (measured from the given point), and find the solution for $\delta$ in leading order. Performing this for $\Gamma$ point,
we find the following results in the case $\alpha> 1$:
\begin{align}
\label{eq:corner-E1}
\frac{\epsilon_1}{t_1}\approx&-1-2\alpha+\frac{\alpha |\vec{k}|^2a^2 }{4 (\alpha +1)},\\
\label{eq:corner-E34}
\frac{\epsilon_{2,3}}{t_1}\approx& 1-\frac{a^2\alpha}{4\left(\alpha ^2-1\right)}\nonumber\\
\times&\bigg[\alpha|\vec{k}|^2\pm \sqrt{\left(\alpha ^2 |\vec{k}|^4-4 (\alpha^2-1) k_x^2 k_y^2\right)}\bigg],\\
\label{eq:corner-E2}
\frac{\epsilon_4}{t_1}\approx&-1+2\alpha+\frac{\alpha |\vec{k}|^2 a^2 }{4 (\alpha-1 )}.
\end{align}
The numbering of bands goes from the lower one to the upper one (for $\alpha<1$ the indices $2$ and $4$ should
be interchanged). From expression \eqref{eq:corner-E1} one can conclude that spectrum of tight-binding Hamiltonian \eqref{eq:tight-binding-Hamiltonian} is bounded by $-1-2\alpha<\epsilon <1+2\alpha$ at zero temperature. In particular, it follows from Eq.(\ref{eq:corner-E34})
that the top of band $\epsilon_3$ has completely flat lines along $k_x$ and $k_y$ axes.

In the case $\alpha=1$ we find the following expansions for three upper bands (which have triply degenerate point (see also Ref.\cite{Yamashita2013PRB})):
\begin{align}\label{eq:corner-alpha-1}
&\frac{\epsilon_1}{t_1}\approx-3+\frac{1}{8} a^2 |\vec{k}|^2, \quad \frac{\epsilon_3}{t_1}\approx 1-\frac{k_x^2 k_y^2 a^2}{2|\vec{k}|^2},\nn
&\frac{\epsilon_{2,4}}{t_1}\approx 1\pm \frac{a}{\sqrt{2}}|\vec{k}|-\frac{a^2 \left(k_x^2-k_y^2\right){}^2}{16 |\vec{k}|^2}.
\end{align}
The two bands $\epsilon_{2,4}$ form Dirac cones with Fermi velocity $v_F={a t_1}/{\sqrt{2}\hbar}$ with additional square-order corrections in $|\vec{k}|a$.
The middle band $\epsilon_3$ is completely flat in first-order approximation, but has nontrivial anisotropic corrections of second-order in $|\vec{k}|a$.

\begin{figure*}
	\centering
	\includegraphics[scale=0.45]{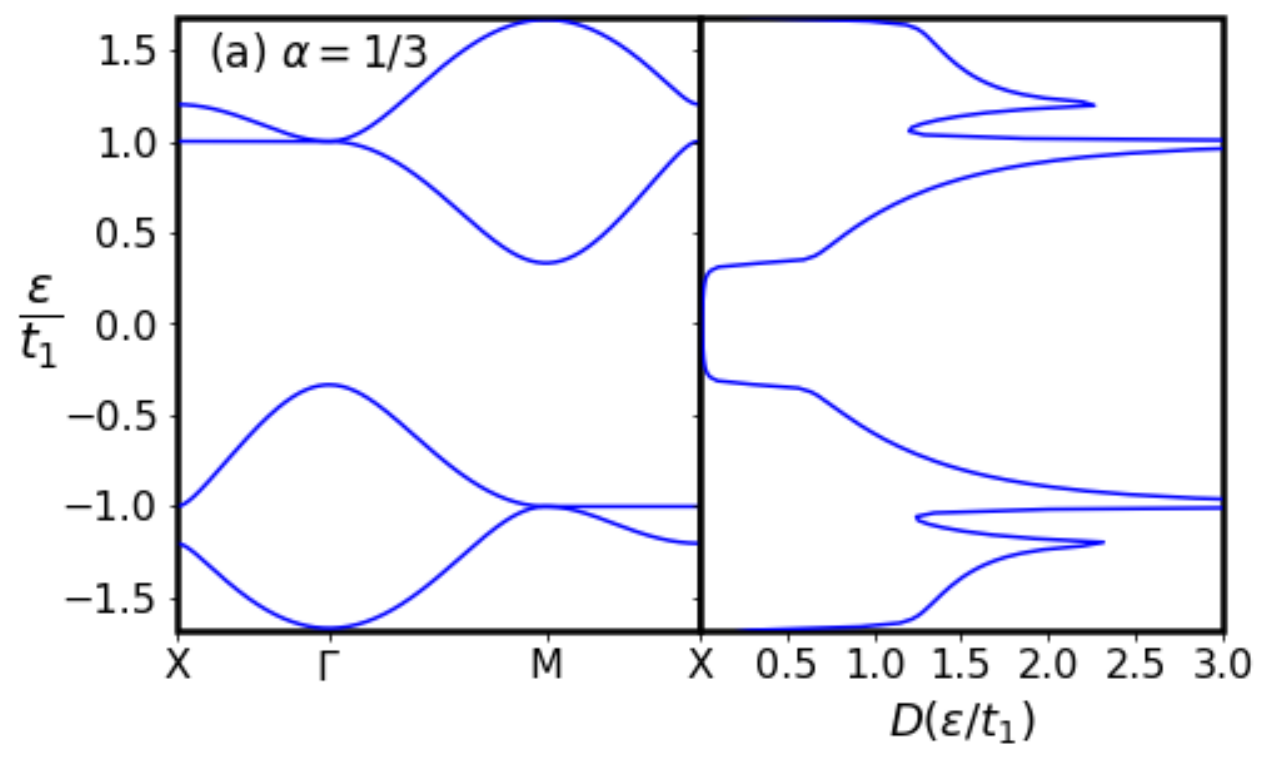}
	\includegraphics[scale=0.45]{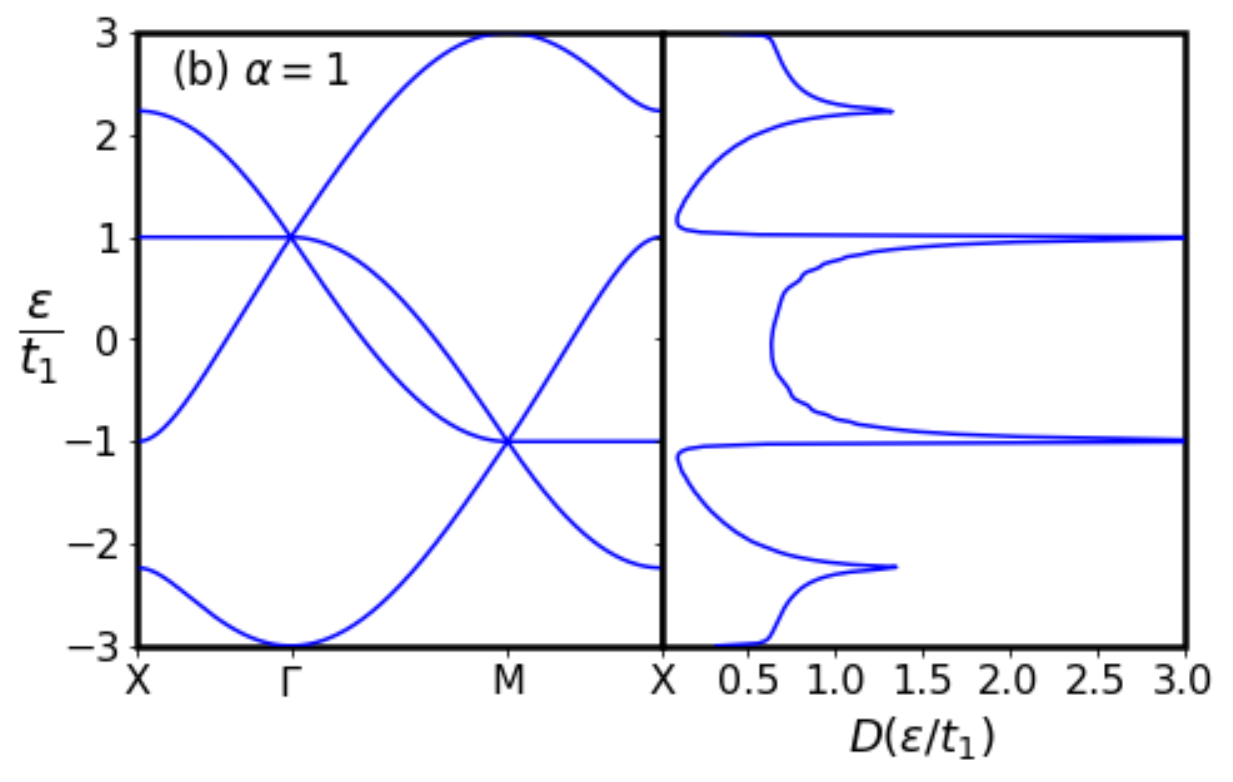}
	\includegraphics[scale=0.45]{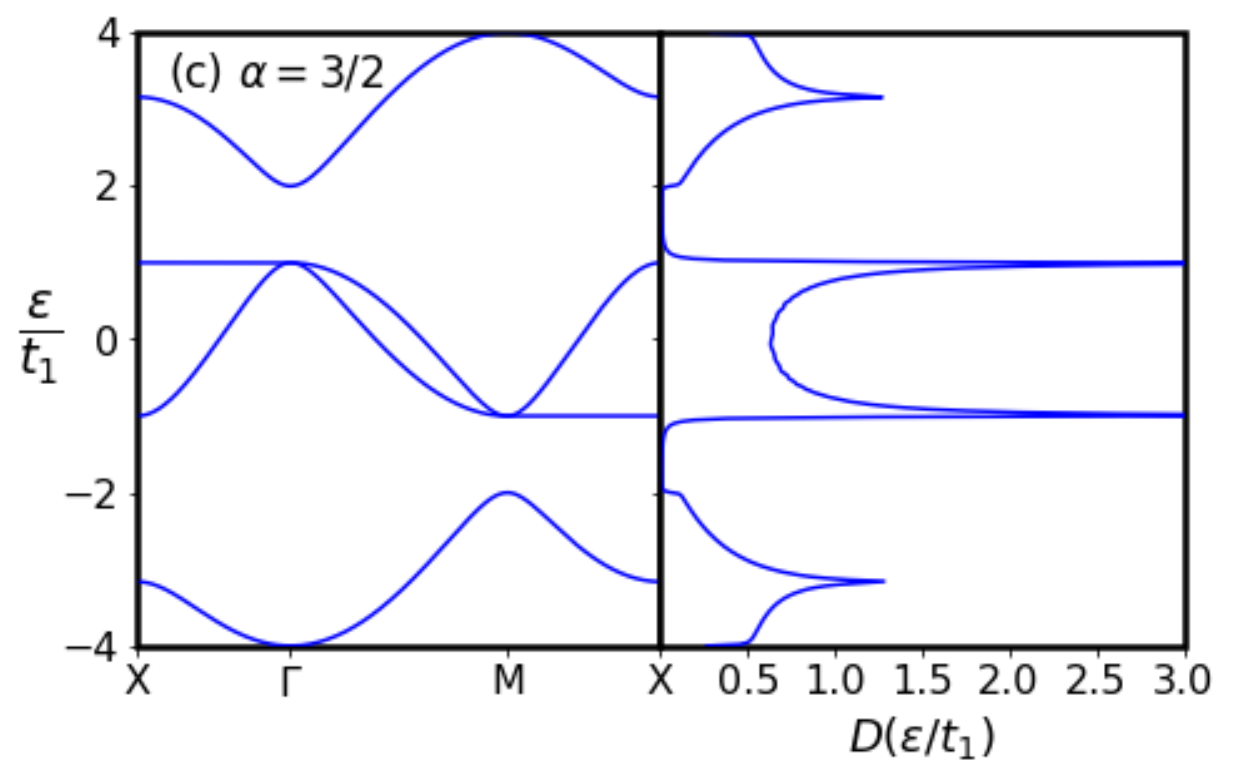}
	\caption{The spectrum of $T$-graphene along  the closed path $X-\Gamma-M-X$ and DOS for $\alpha=1/3,\,1$ and $\alpha=3/2$.  DOS is plotted on the right of each spectrum, and is measured in units of $\frac{1}{a^2 t_1}$.  DOS is regularized with finite broadening of levels, $\Gamma=0.01 t_1$ to make plots smooth.}
	\label{fig:spectrum-dos}
\end{figure*}

The $\Gamma$ and $M$ points define the energy boundaries of each band (see Fig.\ref{fig:spectrum-moved}). For $\alpha\leq 1$ the bands are in the ranges $ [-1-2\alpha,-1],\,[-1,-1+2\alpha],\,[1-2\alpha,1],\,[1,1+2\alpha]$ measured in units of $t_1$. It follows from the expansions \eqref{eq:corner-E1}-\eqref{eq:corner-E2} taken at $\vec{k}=0$. We find that the gap near $\epsilon=0$ opens for $\alpha<1/2$. For the $\alpha\geq 1$
the bands' energy ranges are $\epsilon/t_1\in[-1-2\alpha, 1-2\alpha]$, $[-1,1]$ for both middle bands, and $[-1+2\alpha,1+2\alpha]$. In this case the gaps are
opened for $\alpha>1$ above $\epsilon=t_1$ and below $\epsilon=-t_1$, respectively. These features of spectrum are manifested in vanishing DOS in corresponding
gap energy ranges, see Fig.\ref{fig:spectrum-dos}.

Next, we identify the type of VHS at $\epsilon_{3}=t_1$ in $\alpha=1$ case. For this purpose, we evaluate the DOS contribution for each band separately, taking the leading term in wavevector expansion. The integration over wavevector in Eq.\eqref{eq:dos-definition-int} is extended to cut-off parameter $\Lambda$ of effective expansions \eqref{eq:corner-alpha-1}. Then, the Dirac cones give the standard graphene-like result:
\begin{align}
	D_{2}(\epsilon)+D_{4}(\epsilon)=\frac{|\epsilon-t_1|}{\pi a^2 t_1^2}.
\end{align}
The evaluation of DoS for middle nearly flat band is more complicated, but can be performed in polar coordinates:
\begin{align}
D_{3}(\epsilon \lesssim t_1)=\hspace{-1mm}\int_{0}^{\Lambda}\hspace{-1mm}\int_{0}^{2\pi} \frac{k dk d\phi}{(2\pi)^2}
\delta\hspace{-1mm}\left[\epsilon-t_1+t_1\frac{k^2 a^2 \sin^2(2\phi)}{8}\right].
\end{align}
We emphasize the fact that the middle band contributes only for $\epsilon <t_1$ and the corresponding DOS is asymmetric. The integration over $k$ is easily
performed, and the integration over angle can be confined to first quadrant with adding a total factor $4$. Then, one should integrate in the limits where
the solutions under delta-function are possible:
\begin{align}
&\phi_{min}=\frac{1}{2}\arcsin\left(\sqrt{\frac{8(1-\epsilon/t_1)}{\Lambda^2 a^2}}\right)<\phi<\nn
&\phi_{max}=\frac{\pi}{2}-\frac{1}{2}\arcsin\left(\sqrt{\frac{8(1-\epsilon/t_1)}{\Lambda^2 a^2}}\right).
\end{align}
Thus, the integral for DOS becomes
\begin{align}
D_{3}(\epsilon\lesssim t_1)&=\frac{1}{t_1 a^2}\int_{\phi_{min}}^{\phi_{max}} d\phi\frac{4}{\sin^2 (2\phi)}\nonumber\\
&\approx \frac{2}{t_1 a}\frac{\Lambda }{\sqrt{2(1-\epsilon/t_1)}}.
\end{align}
with the $1/\sqrt{1-\epsilon/t_1}$ divergence, as was noted previously.
This power-law divergence together with asymmetry of the DOS clearly indicates, that this point corresponds to high-order VHS (see middle peaks
of the DOS in all panels of  Fig.\ref{fig:spectrum-dos}). Below we show that this holds true for all points on flat lines in the dispersion $\epsilon_3(\mathbf{k})$. Also one should note that
this singularity has larger exponent $\kappa={1}/{2}$ (which is defined as $D_{3}(\epsilon\leq t_1)\sim |t_1-\epsilon|^{-\kappa}$) than in twisted bilayer graphene ($\kappa={1}/{4}$, \cite{Yuan2019Nature}), and the same as in $\mathrm{Sr}_{3} \mathrm{Ru}_{2} \mathrm{O}_{7}$ \cite{Efremov2019} and $\beta-\mathrm{YbAlB}_{4}$ \cite{Ramires2012PRL} materials.

Above we have found the long wavelength expansions of spectrum for small values of wavevector $\vec{k}$. However, these expansions are violated if the model
parameter $\alpha$ approaches $1$. In this case we can use another series expansion of the spectrum: we assume that $|1-\alpha|\sim |\vec{k}a|$  are of the
same order. Then, we replace both terms $|1-\alpha|$ and $|\vec{k}a|$  in Eq.\eqref{eq:spectral_eq} with  $\zeta|1-\alpha|$ and $\zeta|\vec{k}a|$, respectively, and expand the obtained equation into powers of $\zeta$. This guaranties that expansions keep contributions from both small values $|1-\alpha|$ and $|\vec{k}a|$ in the same leading order. Next, we solve the approximate spectral equation around each band, as for Eqs.\eqref{eq:corner-E1}-\eqref{eq:corner-E2}, and set finally $\zeta=1$, we find
\begin{align}
\label{eq:small-alpha-expansion}
&\frac{\epsilon_{1}}{t_1}=-1-2\alpha+\frac{(k_x^2+k_y^2)a^2}{8},\quad \frac{\epsilon_{3}}{t_1}=1-\frac{k_x^2 k_y^2 a^2}{2(k_x^2+k_y^2)},\nn &\frac{\epsilon_{2,4}}{t_1}=1-\hspace{-1mm}\left(\hspace{-1mm}(1-\alpha)\pm \sqrt{\frac{(k_x^2+k_y^2)a^2}{2}+(1-\alpha)^2}\right).
\end{align}
The last two expressions  show that the $|1-\alpha|$ competes with $|\vec{k}|a$ and their larger value defines the spectrum form in the leading order.

\subsection{$X$-points and flat lines}
At $X$ point the eigenvalues of Hamiltonian \eqref{eq:tight-binding-Hamiltonian} are
\begin{align}
\epsilon_{1,4}^{X}=\mp t_1\sqrt{1+4\alpha^2}, \quad \epsilon_{2,3}^{X}=\mp t_1.
\end{align}
The energies $\epsilon_{1,4}$ belong to lower and upper and bands, respectively, and the energies $\epsilon_{2,3}$ belong to flat lines for the points in k-space, which are situated in the middle between band-touching points. In Appendix \ref{appendix:flat-lines} we show how the flat lines are related to the $C_{4}$ point symmetry group of the lattice and structure of tight-binding Hamiltonian. Performing the series expansion of spectral equation in the same way as discussed above Eq.\eqref{eq:corner-E1} but for wavevectors around $X=(0, \frac{\pi}{a})$, we find:
\begin{align}
\label{Xpoint1}
&\epsilon_{1}\approx \epsilon_{1}^{X}+\frac{t_1 a^2}{4}\bigg[k_x^2 \left(1+\frac{t_1}{\epsilon_{1}^{X}}\right)-\left(k_y-\frac{\pi}{a}\right)^2 \left(1-\frac{t_1}{\epsilon_{1}^{X}}\right)\bigg],\\
\label{Xpoint2}
&\epsilon_{4}\approx\epsilon_{4}^{X}+\frac{t_1 a^2}{4}\bigg[k_x^2\left(1+\frac{t_1}{\epsilon_{4}^{X}}\right)
-\left(k_y-\frac{\pi}{a}\right)^2\left(1-\frac{t_1}{\epsilon_{4}^{X}}\right)\bigg].
\end{align}
These two dispersion relations represent ordinary VHS, defined via the conditions \eqref{eq:van-hove-ordinary}. The Hessian matrix
is diagonal and its' elements are the derivatives of above dispersion relations with respect to wavevectors, $\mathcal{D}=\text{diag}(\p_{xx}
\epsilon, \p_{yy} \epsilon)$. The DOS exhibits a logarithmic divergence around $\epsilon=\epsilon_{1}^{X}$ and $\epsilon=\epsilon_{4}^{X}$:
$D_{1,4}(\epsilon)\sim \log\left(\frac{\Lambda a^2 t_1}{|\epsilon-\epsilon_{1,4}^{X}|}\right)$. These upper and lower peaks in DOS are clearly
visible on Fig.\ref{fig:spectrum-dos}.

Next, we find the series expansion of $\epsilon_{2,3}$ bands' dispersion around $X$ point. Due to chiral symmetry mentioned after Eq.(\ref{eq:spectral_eq}),
it suffices to make expansion only for upper band, while for lower band it can be found by appropriate change of wavevectors. Expanding the spectral
equation \eqref{eq:spectral_eq} for third band around energy $\epsilon_{3}=t_1$ into series in $k_x a$, we find:
\begin{align}\label{eq:X-point-high-order}
	&\epsilon_{3}\approx t_1-t_1\left[\frac{k_x^2 a^2}{2}-\frac{k_x^4a^4}{4\alpha^2(1-\cos(k_y a))}\right].
\end{align}
This approximation works well only for $\frac{k_x^4a^4}{4\alpha^2(1-\cos(k_y a))}<\frac{k_x^2 a^2}{2}$, since this band has $\epsilon_{3}\leq t_1$ energy for
all points in BZ. The Hessian matrix for the dispersion \eqref{eq:X-point-high-order} has only one nonzero component on diagonal $\mathcal{D}=\text{diag}\left(-\frac{t_1 a^2}{2}, 0\right)$. Thus, we observe that the middle bands at X-point and in other points of flat line where
$1-\cos(k_y a)\neq 1$ exhibit a high-order saddle point ($\det {\cal D}=0$).  One can check that the DOS for dispersion \eqref{eq:X-point-high-order}
has a inverse square root divergence $1/\sqrt{t_1-\epsilon}$ with energy, with benchmark asymmetry:
\begin{align}
&D_{3}(\epsilon\lesssim t_1)=\int \frac{d^{2} k}{(2 \pi)^{2}} \delta\left[\varepsilon-t_1\left(1-\frac{k_x^2 a^2}{2}\right)\right]\nn
&=\frac{\Lambda}{\sqrt{2}\pi^2 a t_1 \sqrt{1-\epsilon/t_1}}.
\end{align}
In Fig.\ref{fig:spectrum-dos} we present dispersion relations for $T$-graphene along the path $X-\Gamma-M-X$ which represents the main features in spectrum
(left part of each panel) and DOS (regularized by finite level broadening, right part of each panel) for the values $\alpha=1/3,\,1$ and $\alpha=3/2$.
Note that the path length in $M-\Gamma$ direction is $\sqrt{2}$ times larger than that in $X-M$ or $\Gamma-X$ directions. Our plots show that at
energies $\mp t_{1} \sqrt{1+4 \alpha^{2}}$ DOS exhibits logarithmic divergences, which are the standard VHS at X points. At the same time, the much stronger
peaks in DOS correspond to flat lines in spectrum at energies $\mp t_1$ which are 'high-order' VHS.  Our results for spectra agree with the results of Refs.\cite{Sheng2012,Bao2014Nature,Yamashita2013PRB}, however, the dispersion $\varepsilon_3$ in Eq.(\ref{eq:corner-alpha-1}) was not recognized
as the one exhibiting high-order VHS.

Fig.\ref{fig:spectrum-dos} demonstrates also evolution of DOS as the function of the hopping parameter $\alpha$. At $\epsilon = 0$ we find that for $\alpha<1/2$
there are no states (insulating phase), while for larger $\alpha$ the states are present. For energies $|\epsilon|<t_1$ the DOS is always finite for
$\alpha>1/2$ meaning metallic behavior. On the other hand, for energies $|\epsilon|>t_1$ and $\alpha>1$ we observe the presence of gaps in the DOS.

In Sec.\ref{sec:susceptibility} we will study the behavior of orbital susceptibility around  van Hove singularities.

\subsection{Effective models of band touching point: linear and quadratic approximations}
In the tight-binding model of square-octagon lattice the band touching exists at two highly-symmetric points - $\Gamma$ and $M$. Since they are related by
chiral symmetry (see discussion after Eq.\eqref{eq:spectral_eq}), we need to build an effective Hamiltonian only at one of these points. As was proposed in Ref.\cite{Yamashita2013PRB}, one can perform a rotation to $C_{4v}$ basis utilizing the following unitary matrix
\begin{align}
	\label{eq:C-4v-basis}
	U_{C_{4v}}=\frac{1}{2}\left(\begin{array}{cccc}
		1 & \sqrt{2} & 0 & 1 \\
		1 & 0 & \sqrt{2} & -1 \\
		1 & -\sqrt{2} & 0 & 1 \\
		1 & 0 & -\sqrt{2} & -1
	\end{array}\right),
\end{align}
which acts on four-component wave functions in sublattice space, defined below Eq.\eqref{eq:tight-binding-Hamiltonian}.
After such unitary transformation we obtain the following first-order effective $SU(3)$ Hamiltonian near the $\Gamma$ point:
\begin{align}\label{eq:hamiltonian-Yamashita}
H^{(1)}_{SU(3)}=t_1\left(
\begin{array}{ccc}
1 & 0 & -\frac{i a k_x}{\sqrt{2}}
\\
0 & 1 & \frac{i a k_y}{\sqrt{2}}
\\
\frac{i a k_x}{\sqrt{2}} &
-\frac{i a k_y}{\sqrt{2}} & 2
\alpha -1 \\
\end{array}
\right).
\end{align}
This Hamiltonian is useful for understanding how the Dirac cones emerge in spectrum for $\alpha=1$. The spectrum defined by this Hamiltonian is
\begin{align}
\frac{\varepsilon_{0}}{t_1}=1, \quad \frac{\varepsilon_{\pm}}{t_1}=\alpha \pm \sqrt{\frac{a^{2}|\mathbf{k}|^{2}}{2}+(\alpha-1)^{2}},
\label{spectrum:effective-linearH}
\end{align}
where $\epsilon_0$ corresponds to the $\epsilon_{3}$ band of tight-binding model, and $\epsilon_{-,+}$ to the bands $\epsilon_{2,4}$ respectively.
The corresponding eigenvectors are
\begin{align}
&\Psi_0=	\frac{1}{|\vec{k}|}\left(k_y,k_x,0\right),\nn &\Psi_{-}=\frac{(ik_x a,-ik_y a, \sqrt{2}(1-\epsilon_{-}))}{\sqrt{2\left(|\vec{k}|^2 a^2+2(1-\alpha)(1-\epsilon_{-})\right)}},\nn
&\Psi_{+}=\frac{(-i k_x a,i k_y a, \sqrt{2}(\epsilon_{+}-1))}{\sqrt{2\left(|\vec{k}|^2 a^2+2(\alpha-1)(\epsilon_{+}-1)\right)}}.
\end{align}
One should note that the linear Hamiltonian of such type does not capture the spectral structure of middle band. Instead, the middle band is treated
as completely flat, and the corresponding effective theory is an example of pseudospin-1 fermion models  (see Ref.\cite{Louvet2015} for topological
classification of such theories). Since the aim of present paper is to analyze the role of high-order VHS, we need to build the
effective Hamiltonian that correctly captures the dispersion of middle band at leading order in $|\vec{k}|a$. The needed dispersion is presented, for
example, in Eq.\eqref{eq:corner-alpha-1} in the $\alpha=1$ case.

To find corresponding effective Hamiltonian, we use L\"{o}wdin method \cite{Lowdin1951}, which is also called L\"{o}wdin partitioning (the example
calculation for Lieb-kagome Hamiltonian was performed in Ref.\cite{Lim2020PRB}). The idea is to perform the rotation of the full tight-binding Hamiltonian \eqref{eq:tight-binding-Hamiltonian} via the unitary transformation \eqref{eq:C-4v-basis}, and then represent it in a block-like form:
\begin{align}
H=\left(\begin{array}{c|c}
{H_{\alpha \alpha}} & {H_{\alpha \beta}} \\
\hline H_{\beta \alpha} & {H_{\beta \beta}}
\end{array}\right),
\end{align}
where the $\alpha$ subspace describes $SU(3)$ band-touching and $\beta$ subspace corresponds to lower band, decoupled from other three bands by relatively large gap. Then, the effective second-order Hamiltonian around band-touching is written as
\begin{align}\label{eq:Lodwin-formula}
\mathcal{H}_{\alpha}=H_{\alpha \alpha}+H_{\alpha \beta}\left(\epsilon_{0}-H_{\beta \beta}\right)^{-1} H_{\beta \alpha},
\end{align}
where $\epsilon_{0}=\epsilon_{2,3}(\vec{k}=0)=t_1$. For $\Gamma$ point this Hamiltonian has the following form
\begin{align}\label{eq:Yamashita-Lodwin-a-all}
H_{SU(3)}^{(2)}=\hat{\epsilon}^{(0)}+t_1\left(
\begin{array}{ccc}
-\frac{a^2 (2 \alpha +1) k_x^2}{4 (\alpha +1)} &
\frac{a^2 k_x k_y}{4 (\alpha +1)} & -\frac{i a
	k_x}{\sqrt{2}} \\
\frac{a^2 k_x k_y}{4 (\alpha +1)} & -\frac{a^2 (2 \alpha
	+1) k_y^2}{4 (\alpha +1)} & \frac{i a k_y}{\sqrt{2}} \\
\frac{i a k_x}{\sqrt{2}} & -\frac{i a k_y}{\sqrt{2}} &
\frac{\vec{k}^2 a^2}{4} \\
\end{array}
\right),
\end{align}
where $\hat{\epsilon}^{(0)}=t_1\text{diag}(1,\,1,\,2 \alpha -1)$ . Such simple Hamiltonian is particularly useful when the proper dispersion of all three bands is needed at leading order.

\section{Orbital susceptibility}
\label{sec:susceptibility}
In this section we study the manifestation of T-graphene spectrum features considered above, in particular, VHS of both kinds, in the orbital susceptibility.
The susceptibility  measures the response of a electronic system to an external magnetic field and is defined standardly as the second derivative
of the grand canonical potential at zero field.
The main formula, which is most suitable in our case for numerical calculation, was given in Ref.\cite{Piechon2016PRB},  the more general formula was derived in Ref.\cite{Raoux2015PRB}. The susceptibility can be
represented as
\begin{align}\label{eq:susceptibility}
\chi_{\text {orb }}(\mu, T)=-\frac{\mu_{0} e^{2}}{12 \hbar^{2}} \frac{\Im}{\pi S} \int_{-\infty}^{\infty} n_{\mathrm{F}}(\epsilon) \operatorname{Tr} \hat{X} \mathrm{d} \epsilon.
\end{align}
Here $n_F(\epsilon)=1 /(e^{(\epsilon-\mu) / T}+1)$ is the Fermi distribution, $\mu_{0}=4 \pi \times 10^{-7}$ in SI units and S is the area of the sample.
The operator $\hat{X}$ is written in terms of zero-field Green function $G(\vec{k})$ and Bloch Hamiltonian $H(\vec{k})$, and $\p_{x,y}$ are partial
derivatives over momenta:
\begin{align}\label{eq:susceptibility-X-operator}
\hat{X}&=G(\vec{k}) \partial_{x}^{2} H(\vec{k}) G(\vec{k}) \partial_{y}^{2} H(\vec{k})-\nn
&-G(\vec{k}) \partial_{x y}^{2} H(\vec{k}) G(\vec{k}) \partial_{x y}^{2} H(\vec{k})+\nn
&+2\left(\left[G(\vec{k}) \partial_{x} H(\vec{k}), G(\vec{k}) \partial_{y} H(\vec{k})\right]\right)^{2}.
\end{align}
The trace operation contains the integral over the BZ and the trace over band indices:
\begin{align}\label{eq:trace-operation-definition}
	\operatorname{Tr}(\bullet)=\sum_{k} \operatorname{tr}(\bullet)=S \int_{\mathrm{BZ}} \frac{\mathrm{d}^{2} k}{4 \pi^{2}} \operatorname{tr}(\bullet).
\end{align}
The orbital susceptibility can be rewritten in several other forms, one of them without commutator  \cite{Raoux2015PRB},
\begin{align}\label{eq:24-in-Raoux2015}
\chi_{\text {orb }}(\mu, T)=&-\frac{\mu_{0} e^{2}}{12 \hbar^{2}} \frac{\operatorname{Im}}{\pi S} \int_{-\infty}^{+\infty} n_{F}(\epsilon)
\operatorname{Tr}\left\{G H^{x x} G H^{y y}\right.\nn
&-\left.G H^{x y} G H^{x y}-4\left(G H^{x} G H^{x} G H^{y} G H^{y}\right.\right.\nonumber\\
&-\left.\left. G H^{x} G H^{y} G H^{x} G H^{y}\right)\right\} \mathrm{d} \epsilon.
\end{align}
Here $G=G(\vec{k})$ is the Green function and $H^{i},\,H^{ij}$ denote the first and second derivatives of Hamiltonian with respect to components
of momenta $k_{i,j}$ and the trace contains momenta integration, as defined in Eq. \eqref{eq:trace-operation-definition}. The last formula can be
also rewritten \cite{Raoux2015PRB} in terms of previously found one by Gomez-Santos \cite{Gomez-Santos2011PRL},
\begin{align}\label{eq:26-in-Raoux2015}
&\chi_{\mathrm{orb}}(\mu, T)=-\frac{\mu_{0} e^{2}}{2 \hbar^{2}} \frac{\mathrm{Im}}{\pi S} \int_{-\infty}^{+\infty} n_{F}(\epsilon)
\operatorname{Tr}\left\{G H^{x} G H^{y}G H^{x}\right.\nn
&\times\left.  G H^{y}+\frac{1}{2}\left(G H^{x} G H^{y}+G H^{y} G H^{x}\right) G H^{x y}\right\} \mathrm{d} \epsilon.
\end{align}
Here the first term represents the Fukuyama result \cite{Fukuyama1971}. Three formulas for susceptibility are equivalent of course, and the use of a
specific formula depends on possible simplifications, for example, for Hamiltonians linear in momenta the expressions (\ref{eq:susceptibility-X-operator})
or (\ref{eq:24-in-Raoux2015}) are preferred since the terms with second derivatives $H^{ij}$ vanish.

To check the numerical results below we use the sum rule which states that the integral of the orbital susceptibility over the whole
band vanishes:
\begin{align}\label{eq:sum-rule}
\int \chi_{\mathrm{orb}}(\mu, T) d \mu=0.
\end{align}
The derivation of the sum rule for general tight-binding model was given in Ref.\cite{Raoux2014}. Below we apply the formulas for orbital susceptibility to
particular models, namely - tight-binding model of tetragraphene and effective low-energy $SU(3)$ models.

\subsection{Application of general formulas to tetragraphene}

Let us now apply the formula \eqref{eq:susceptibility} to tetragraphene Hamiltonian (\ref{eq:tight-binding-Hamiltonian}). Since the second derivatives
$\p_{xy}^2H$ and $\p_{yx}^{2}H$ vanish,  the operator $\hat{X}$ reduces to
\begin{align}
&\hat{X}=G(\vec{k}) \partial_{x}^{2} H(\vec{k}) G(\vec{k}) \partial_{y}^{2} H(\vec{k})+\nn
&+2\left(\left[G(\vec{k}) \partial_{x} H(\vec{k}), G(\vec{k}) \partial_{y} H(\vec{k})\right]\right)^{2}.
\label{X-matrix}
\end{align}
The Green's function  is given in Appendix \ref{sec:green-func}. Then, calculating the trace of $\hat{X}$ for each term separately, we find the expressions presented in Appendix by Eqs.\eqref{eq:term1-polynomial} and \eqref{eq:term2-polynomial}.
We denote the first term with second derivatives in (\ref{X-matrix}) as ``\textit{term 1}'' and the term with commutator as ``\text{term 2}''.
Here and thereafter we use dimensionless energy parameter $\epsilon\to \epsilon/t_1$  to simplify the form of expressions. One should notice that the numerators in both terms \eqref{eq:term1-polynomial} and \eqref{eq:term2-polynomial} are real, thus the
imaginary part comes fully from integration over energy due to the presence of singular denominators.  We write the determinants as
$\prod\limits_{i=1}^{4}(\epsilon -\epsilon_{i}(\vec{k}))$, where $\epsilon_{i}(\vec{k})$ are band energies measured in units of $t_1$.

One can use also an alternative expression \eqref{eq:26-in-Raoux2015} for susceptibility obtaining shorter expression
\begin{eqnarray}
\label{eq:reduced-26-Raoux-suscep}
&&\chi_{\mathrm{orb}}(\mu, T)=-\frac{\mu_{0} e^{2} t_1}{2 \hbar^{2}}\frac{\mathrm{Im}}{\pi } \int\limits_{-\infty}^{+\infty}d\epsilon n_{F}(t_1\epsilon)\nn
&&\times \int\limits_{BZ}\frac{d^2 k}{4\pi^2} \operatorname{tr}\left\{G H^{x} G H^{y} G H^{x} G H^{y}\right\}.
\end{eqnarray}
Evaluating the trace, we find
\begin{align}
\label{eq:trace-T-graphene-simple}
&\operatorname{tr}\left\{G H^{x} G H^{y} G H^{x} G H^{y}\right\}\nn
&=\left(\frac{2\alpha a (\epsilon^2-1)}{\operatorname{det}[\epsilon-\frac{1}{t_1}H(\vec{k})]}\right)^4\sin^{2}(k_x a)\sin^2(k_y a).	
\end{align}
The advantage of this formula is that the numerator is much simpler comparing to Eqs.\eqref{eq:term1-polynomial}-\eqref{eq:term2-polynomial}. However,
the larger power of denominator makes it harder to perform numerical calculation, since the behavior at band-touching point is more singular.

The integrals over energy can be evaluated analytically using Cauchy formula with residues. Next, we need to calculate the integrals over wavevector
in full BZ. They are cumbersome and can be performed only numerically.

The numerical evaluation can be performed by sampling many points in BZ, and replacing integral by a quadrature sum.  For this purpose we use Monte Carlo
approach - it converges very fast with increasing number of sample points for multidimensional integrals. Taking $N$ sample points in BZ, the integral
over $d^2k$ is replaced by the sum $\int_{BZ} \frac{d^2 k}{(2\pi)^2}f(\vec{k})=\frac{1}{N}\sum_{j}f(\vec{k}_j)$.
Then, the final formula used in evaluation is
\begin{align}\label{eq:monte-carlo-integration}
&\chi_{\text {orb }}(\mu, T)=\frac{\chi_0}{N}\sum_{j=1}^{N}\left[\sum_{i}\underset{\epsilon=\epsilon_i}{\text{res}}n_F(t_1 \epsilon)f^{R}(\epsilon)\right]_{\vec{k}=\vec{k}_j}.
\end{align}
The residues were evaluated analytically using expressions \eqref{eq:term1-polynomial}-\eqref{eq:term2-polynomial},  and the band energy solutions
 of spectral equation \eqref{eq:spectral_eq} were substituted numerically into final expressions.
Here we introduced the scale factor for susceptibility $\chi_0=\mu_0 e^2 a^2 t_1/12\hbar^2.$

\begin{figure*}
	\centering
	\includegraphics[scale=0.25]{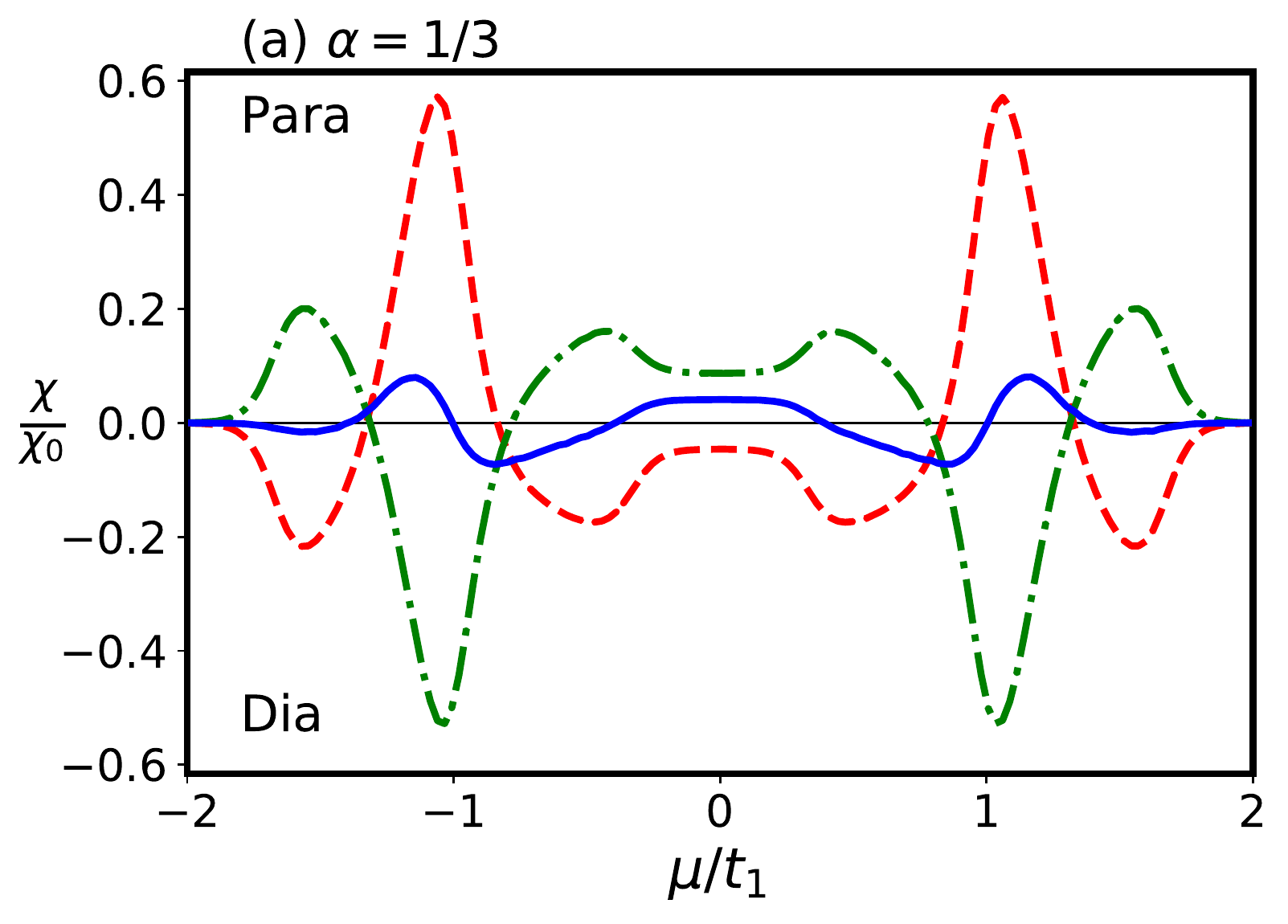}\quad
	\includegraphics[scale=0.25]{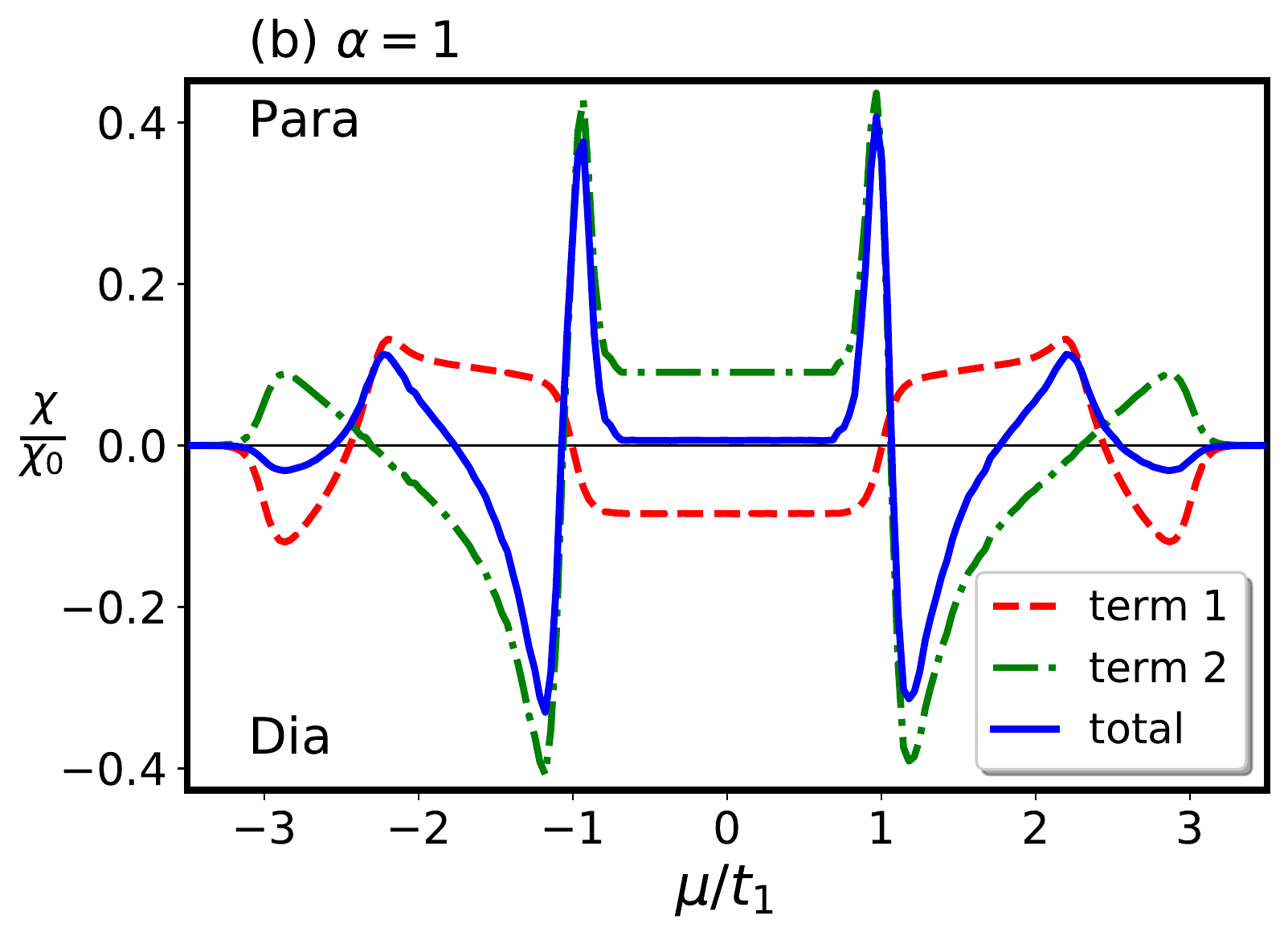}\quad
	\includegraphics[scale=0.25]{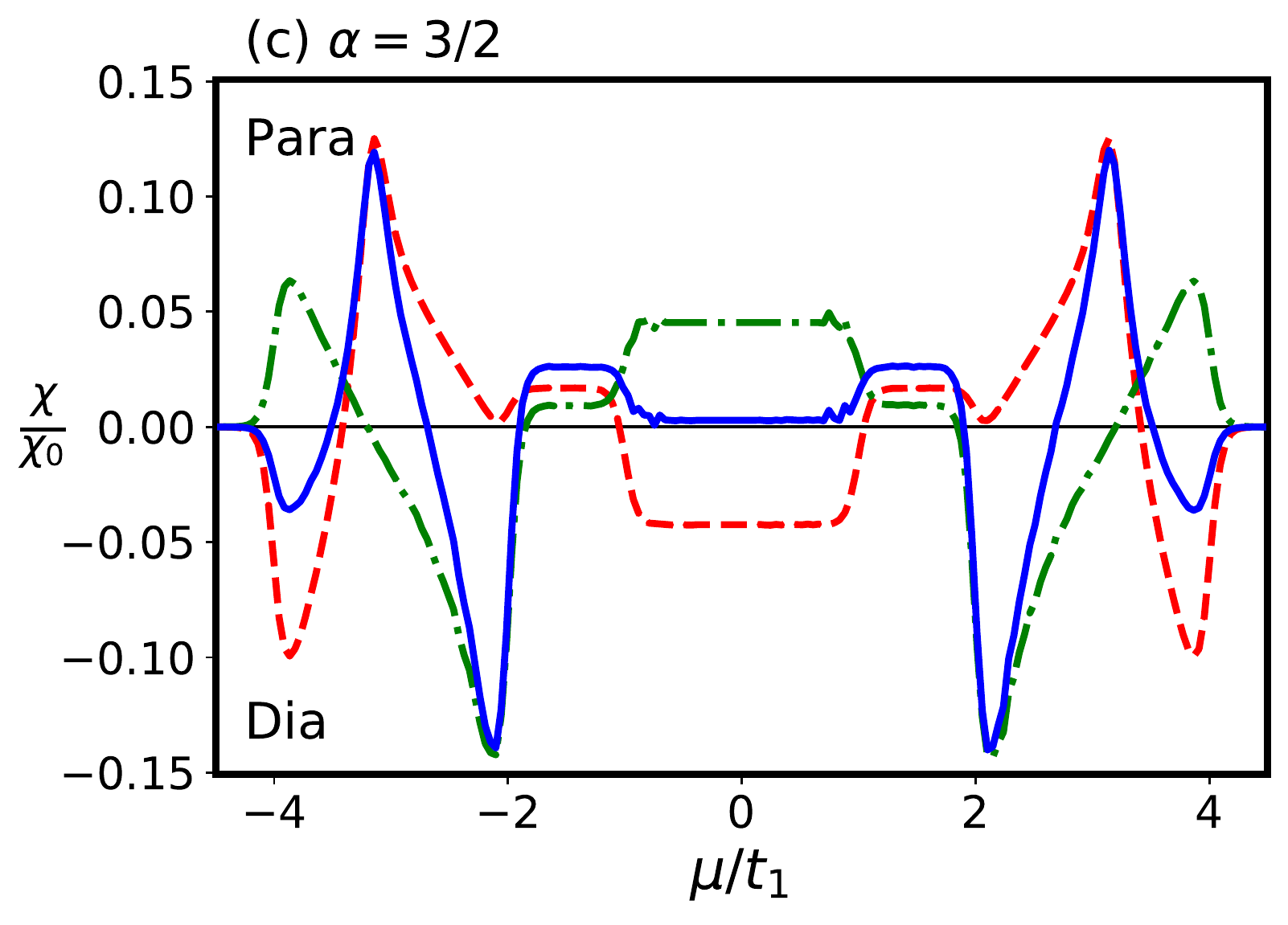}
    \includegraphics[scale=0.25]{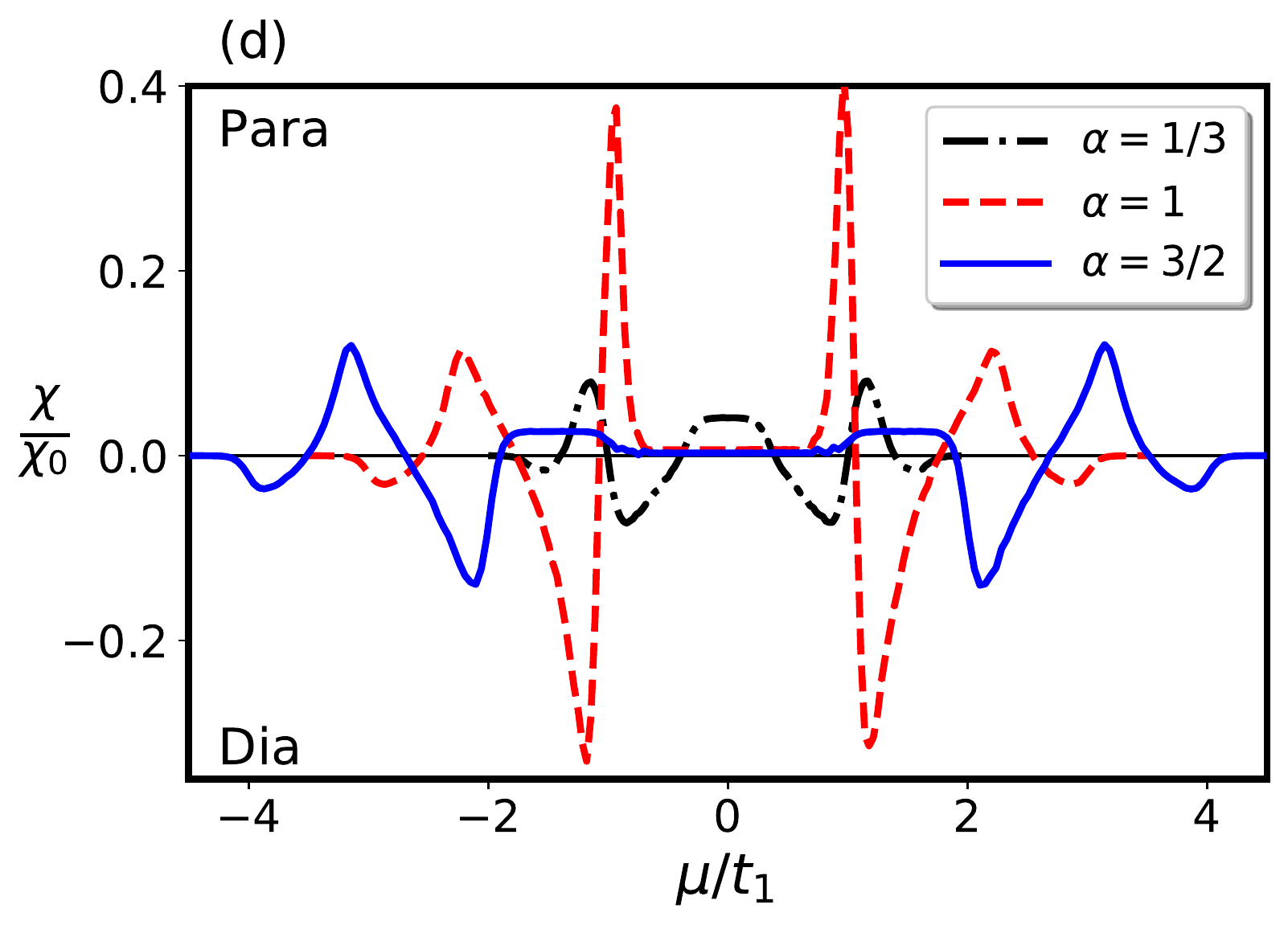}
	\caption{The dependence of susceptibility $\chi$ on chemical potential $\mu$, measured in units of $t_1$ hopping parameter, for three values of $\alpha$: a) $1/3$, b) $1$, c) $3/2$. The susceptibility is normalized to scale factor $\chi_{0}=\mu_{0} e^{2} a^{2} t_{1} / 12 \hbar^{2}$. The legend on  panel (b) shows the lines definitions in  panels a) - c): dashed and dash-dotted lines correspond to first and second term contributions in $\hat{X}$ (see Eq.\eqref{X-matrix}), while the solid line describes the total susceptibility (the different ranges in $y$-axis are taken for better visibility).  Panel d) shows the total susceptibility for three values of $\alpha$. }
	\label{fig:susceptibility-numerical}
\end{figure*}

The results of evaluation for $\chi$ as a function of chemical potential are shown in Fig.\ref{fig:susceptibility-numerical}. We have checked that
good convergence is reached for $N=10^5$ and $N=5\times 10^5$ for the terms \eqref{eq:term1-polynomial} and \eqref{eq:term2-polynomial}, respectively. The errors of integration become in this
case several orders less than the absolute values of susceptibility. As a test, we checked that the sum rule, which is given by Eq.\eqref{eq:sum-rule},
holds true with the same precision.

The orbital susceptibility exhibits standard weak diamagnetic peaks near the edges of the spectrum, which can be easily understood from the Landau-Peierls (LP)
formula \cite{Landau1930, Peierls1933, Raoux2015PRB,Piechon2016PRB},
\begin{align}\label{eq:Landau-Peierls}
&\chi_{\mathrm{LP}}(\mu, T)=\nn
&\frac{\mu_0 e^2}{12\hbar^2}\sum_{i=1}^{4}\int \frac{d^2 k}{4\pi^2} n_F^{\prime}(\epsilon_{i})\left(\partial_{x}^{2} \varepsilon_{i} \partial_{y}^{2} \varepsilon_{i}-\partial_{x y}^{2} \varepsilon_{i} \partial_{x y}^{2} \varepsilon_{i}\right),
\end{align}
which takes into account only intraband contributions. Here $n_F^{\prime}(\epsilon)$ is a derivative of the Fermi distribution function.
We note that the LP contribution in total susceptibility comes from the first two terms in Eq.(\ref{eq:susceptibility-X-operator}) which contain 
second derivatives.

In the case of T-graphene only the lower (upper) band gives strong
contribution to the orbital susceptibility at the lower (upper) edge of the spectrum. This can be clearly seen from Figs.\ref{fig:spectrum-moved} and \ref{fig:spectrum-dos}, since at lower (upper) edge the corresponding band in $\Gamma$ (M) point is separated by a large gap from other three bands.
The dispersion of this band is  quadratic in momenta, see Eq.\eqref{eq:corner-E1}, and both derivatives in first term of LP formula are positive.
The second term exactly vanishes, and thus the LP susceptibility is negative because $n_F^{\prime}(\epsilon)<0$. These peaks are clearly visible
in susceptibility described by the red line (term 1) in panels a) - c) of Fig.\ref{fig:susceptibility-numerical} (leftmost and rightmost
negative peaks).
At the same time, the Landau-Peierls formula does not capture the contribution of high-order saddle points. This is because the large contribution from
a Fermi function derivative $n_{F}^{'}(\epsilon_i)$ is compensated by vanishing determinant of Hessian matrix that is present in round brackets.

At the ordinary van Hove points, which are placed on upper and lower bands at X-points at the energy levels $\epsilon_{1,4}^{X}=\mp \sqrt{1+4\alpha^2}$,
one finds strong paramagnetic peaks. These peaks are also well-described by the Landau-Peierls formula \eqref{eq:Landau-Peierls}. Substituting series
expansion \eqref{Xpoint1} or \eqref{Xpoint2}, one finds that only the first term in Landau-Peierls formula is nonzero, and have positive sign due to
opposite signs
of $\p_x^2$ and $\p_y^2$ derivatives. Moreover, due to the divergent DOS at this energy level, the contribution of this band dominates and leads to
strong paramagnetism. This is also related to famous magnetic breakdown phenomena \cite{Landau-course9}, where the quasiclassical approximation in terms
of electronic orbits fails in the vicinity of saddle points due to effects of tunneling from one trajectory to the neighboring one that leads to rotation
of the electron in a direction opposite to the direction of classical rotation (see Ref.\cite{Vignale1991PRL} for physical picture of this phenomenon).
Large paramagnetic peaks coming from the Landau-Peierls formula are well seen in the red line (term 1) in the left panel of Fig.\ref{fig:susceptibility-numerical} ($\alpha=1/3$). Due to the sum rule (\ref{eq:sum-rule}) they are almost compensated by diamagnetic contribution
in the green line (term 2). The competition of two terms in Eq.(\ref{X-matrix}) leads to several
dia- to paramagnetic transitions when we continuously  change the chemical potential $\mu$ (see Fig.\ref{fig:susceptibility-numerical}). The
susceptibility for $\alpha=3/2$ behaves qualitatively similar to the case with $\alpha=1/3$.

The behavior of the susceptibility is more interesting when the hopping parameter $\alpha$ is close to unity.
At the Fermi level $\mu=0$ the orbital susceptibility does not exhibit any peculiar properties. However, when the doping is tuned to band-touching point
$\mu=t_1$, one can expect nontrivial behavior of susceptibility due to presence of massless fermions forming a Dirac cone and flat lines with high-order VHS
of DOS. Near the energy levels $\mu=\pm t_1$ (see the panels b) and d) in  Fig.\ref{fig:susceptibility-numerical}) we find strong diamagnetic and paramagnetic
peaks. Since the contribution of high-order VHS is suppressed
in the LP formula (term 1) we are left with diamagnetic contribution from the term 2 due to Dirac excitations when $|\mu|\gtrapprox t_1$. On the other
hand, when $|\mu|\lessapprox t_1$ there is a strong paramagnetic contribution in the term 2 from high-order VHS.
The existence of the orbital paramagnetism is a necessary condition to  cancel the diamagnetic contribution in order to satisfy the sum rule (\ref{eq:sum-rule}).
The competition of these two contributions leads to a sharp dia- to paramagnetic transition at $|\mu|\approx t_1$ (see panels b), d) in Fig.\ref{fig:susceptibility-numerical} and Supplement \cite{Supplement}). This transition manifests itself in Fig.\ref{fig:plots-effective1} where the susceptibility at $\mu=t_1$ is plotted
as a function of $\alpha$ (blue line). 

Below we analyze the orbital susceptibility for effective linear and quadratic Hamiltonians given by Eqs.(\ref{eq:hamiltonian-Yamashita}) and (\ref{eq:Yamashita-Lodwin-a-all}) to obtain some insights into the physics of these peculiar features.

\subsection{Analytical results in effective pseudospin-1 model around band-touching}
\label{sec:susceptibility-analytic-Yamashita}

Let us firstly use the linear effective Hamiltonian around band-touching point to find an  analytical approximation for the susceptibility. It is given by Eq.\eqref{eq:hamiltonian-Yamashita}, and we omit the dimensional parameter $t_1$, restoring it in the final expressions for susceptibility,
\begin{align}
H_{3}\equiv \frac{H_{SU(3)}}{t_1}=\left(
\begin{array}{ccc}
1 & 0 & -\frac{i a k_x}{\sqrt{2}}
\\
0 & 1 & \frac{i a k_y}{\sqrt{2}}
\\
\frac{i a k_x}{\sqrt{2}} &
-\frac{i a k_y}{\sqrt{2}} & 2
\alpha -1 \\
\end{array}
\right).
\end{align}
The corresponding Green's function is
\begin{widetext}
\begin{align}
G_{SU(3)}=\frac{1}{\det[\epsilon-H_{3}]}\left(
\begin{array}{ccc}
\epsilon^2-\frac{1}{2} a^2 k_y^2-2 \alpha  (\epsilon-1)-1 & -\frac{1}{2} a^2 k_x k_y & -\frac{i a (\epsilon-1)
	k_x}{\sqrt{2}} \\
-\frac{1}{2} a^2 k_x k_y & \epsilon^2-\frac{1}{2} a^2 k_x^2-2 \alpha  (\epsilon-1)-1 & \frac{i a (\epsilon-1)
	k_y}{\sqrt{2}} \\
\frac{i a (\epsilon-1) k_x}{\sqrt{2}} & -\frac{i a (\epsilon-1) k_y}{\sqrt{2}} & (\epsilon-1)^2 \\
\end{array}
\right).
\end{align}
\end{widetext}

The determinant in denominator is simple
\begin{align}
\hspace{-2mm}\det[\epsilon-H_{3}]=\frac{1-\epsilon}{2} \left(a^2 \vec{k}^2+2 (\epsilon-1) (2 \alpha -\epsilon -1)\right)
\end{align}
and gives two Dirac cones and the flat band at $\varepsilon=1$.
The first derivatives of Hamiltonian are,
\begin{align}
H^{x}_{3}=\frac{a}{\sqrt{2}} \left(
\begin{array}{ccc}
0 & 0 & -i \\
0 & 0 & 0 \\
i & 0 & 0 \\
\end{array}
\right),\,\, H^{y}_{3}=\frac{a }{\sqrt{2}} \left(
\begin{array}{ccc}
0 & 0 & 0 \\
0 & 0 & i \\
0 & -i & 0 \\
\end{array}
\right),
\end{align}
while all second derivatives are zero. Then, we can apply the formula \eqref{eq:26-in-Raoux2015}, which in our case reduces to
\begin{align}
&\chi_{\mathrm{orb}}(\mu, T)
=-\frac{\mu_{0} e^{2}t_1}{2 \hbar^{2}} \times\nn
&\frac{\mathrm{Im}}{\pi S} \int\limits_{-\infty}^{+\infty} n_{F}(\epsilon) \operatorname{Tr}\left\{G H^{x} G H^{y} G H^{x} G H^{y}\right\} \mathrm{d} \epsilon.
\end{align}
Calculating the matrix trace we come at the orbital susceptibility given by the triple integral,
\begin{align}
&\chi_{\mathrm{orb}}(\mu, T)=-\frac{\mu_{0} e^{2} t_1}{2 \hbar^{2}} \frac{\mathrm{Im}}{\pi} \int_{-\infty}^{+\infty} n_{F}(t_1 \epsilon)d\epsilon \,\nn\times
&\int \frac{d^2 k}{4\pi^2} \frac{16 a^8 k_x^2 k_y^2}{\left(a^2
	\left(k_x^2+k_y^2\right)+2 (\epsilon
	-1) (2 \alpha -\epsilon
	-1) \right)^4}.
\end{align}
The integration over momenta is easily performed using polar coordinates
\begin{align}
&\int \frac{d^2 k}{4\pi^2} \frac{16 a^8 k_x^2 k_y^2}{\left(a^2
	\left(k_x^2+k_y^2\right)+2 (\epsilon-1) (2 \alpha -\epsilon-1) \right)^4}\nn
&=\frac{a^2}{12\pi}\times\begin{cases}
\frac{1}{2(\alpha-1)}\left(\frac{1}{\epsilon-1}-\frac{1}{\epsilon+1-2\alpha}\right),&\alpha\neq 1,\\
-\frac{1} {(\epsilon-1)^2},&\alpha=1.
\end{cases}
\end{align}
Then,  using the formula
\begin{align}\label{eq:energy-integral-raouxA6}
\operatorname{Im} \int_{-\infty}^{+\infty} \frac{f(E)}{(E-\alpha)^{j}} \mathrm{d} E=-\frac{\pi}{(j-1) !} f^{(j-1)}(\alpha),
\end{align}
for susceptibility  we finally obtain:
\begin{align}\label{eq:effective-susceptibility}
&\chi_{\mathrm{orb}}(\mu, T)=-\frac{\chi_0}{2\pi}\nn
&\times \begin{cases}
\frac{1}{2(\alpha-1)}\left(n_F(t_1(2\alpha-1))-n_F(t_1)\right),& \alpha\neq 1,\\
t_1n_F^{\prime}(t_1), &\alpha=1.
\end{cases}
\end{align}
Note that the case $\alpha=1$ is the limit of the upper case with $\alpha\neq 1$. The result for $\alpha=1$ has the same functional structure as the
susceptibility for low-energy model of  graphene \cite{Raoux2015PRB}, but differs in numerical factor and sign. The latter difference is connected with
the presence of flat band in spectrum. In such a case the flat band plays the crucial role giving strong delta-like paramagnetic response of the system at
$\mu=t_1$ instead of diamagnetic, which was a result of two Dirac cones in graphene.  Note however, that the linear effective Hamiltonian does not capture
the correct dispersion of the middle band. The model contains completely flat band and the spectrum (\ref{spectrum:effective-linearH}) is similar
to a gapped dice model where the paramagnetic contribution from flat band exceeds diamagnetic contribution from Dirac cones (see Ref.\cite{Piechon2015JPCM})

The plot of effective susceptibility defined by Eq.\eqref{eq:effective-susceptibility} is shown in Fig.\ref{fig:plots-effective1} as a function of
a hopping parameter $\alpha$. On the plot it is denoted as ``Eq.(24)'' effective theory. We compare its dependence on $\alpha$ with total susceptibility
of actual model evaluated numerically. The doping level $\mu=t_1$ coincides with the band touching point at which the high-order VHS and Dirac point are present for $\alpha=1$.
The numerical calculations demonstrate the presence of dia- to paramagnetic transition at $\alpha\approx 0.94$, which is absent in the low-energy result \eqref{eq:effective-susceptibility}. Thus, we should analyze more precise effective model, which is given by the second-order Hamiltonian Eq.\eqref{eq:Yamashita-Lodwin-a-all}.

\subsection{Paramagnetic-diamagnetic phase transition at band-touching point and second-order effective Hamiltonian}
\label{sec:4C}
The calculation of orbital susceptibility for the second-order effective Hamiltonian \eqref{eq:Yamashita-Lodwin-a-all} involves all terms in $\hat{X}$
operator \eqref{eq:susceptibility-X-operator}, because all first and second derivatives of Hamiltonian \eqref{eq:Yamashita-Lodwin-a-all} over $k_{i}$ are nonzero.
The corresponding Green's function is presented in Appendix, see Eq.\eqref{eq:green-lodwin}. Since the calculations quickly become cumbersome, we present only numerical results here. For the integrals over wave number $\vec{k}$ we use Monte-Carlo method. The energies for each point in k-space are found from Eq.\eqref{eq:lodwin-det} and then we use the integration formula \eqref{eq:monte-carlo-integration} multiplied by volume factor $\Lambda^2 a^2/\pi^2$. Here
$\Lambda$ is a cut-off parameter, that defines the region of applicability of second-order effective Hamiltonian \eqref{eq:Yamashita-Lodwin-a-all}. We estimated it as $\Lambda\approx 0.8 \frac{1}{a}$ by comparing
exact spectrum with one obtained from Eq.\eqref{eq:lodwin-det}.

\begin{figure}
	\centering
	\includegraphics[scale=0.5]{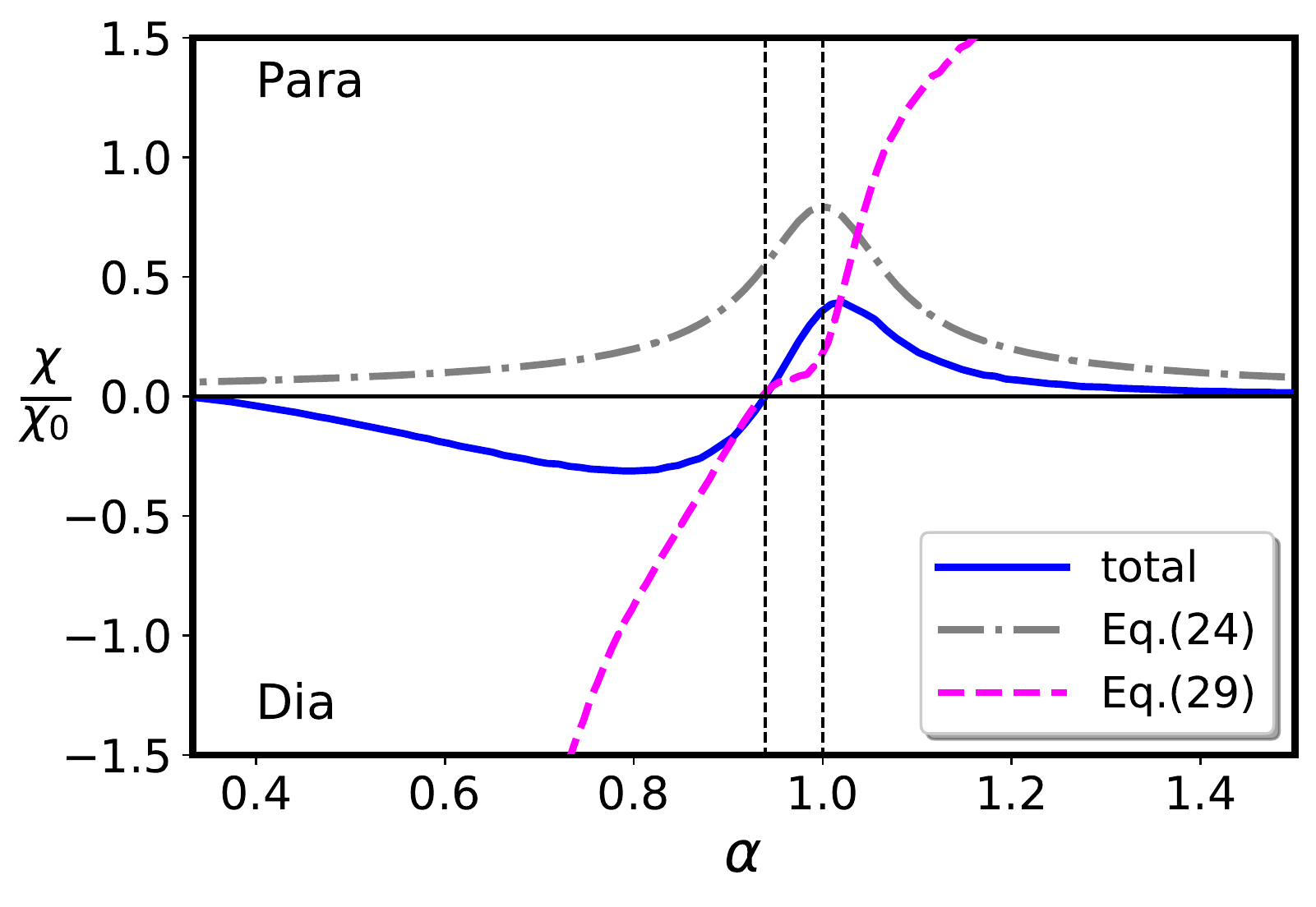}
	\caption{The dependence of orbital susceptibility on relative strength of tight-binding parameters $\alpha=t_2/t_1$ for $\mu=1.0t_1$ and $T=0.05t_1$.
The numerically-evaluated total susceptibility (solid blue line) is compared with susceptibility obtained from effective pseudospin-1 Hamiltonians \eqref{eq:hamiltonian-Yamashita} (gray dash-dotted line) and \eqref{eq:Yamashita-Lodwin-a-all} (magenta dashed line). }
	\label{fig:plots-effective1}
\end{figure}

The orbital susceptibility for the effective Hamiltonian \eqref{eq:Yamashita-Lodwin-a-all} at the band-touching point $\mu=1.0t_1$ as a function of
a hopping parameter $\alpha$ is presented in Fig.\ref{fig:plots-effective1}. It is
clearly seen that this Hamiltonian exhibits dia- to paramagnetic transition at $\alpha=0.94$ in agreement with tight-binding Hamiltonian and in contrast to the
linear effective Hamiltonian \eqref{eq:hamiltonian-Yamashita}.
Qualitatively, one can expect that such a transition occurs due to the presence of Dirac cones, which give strong diamagnetism in graphene \cite{McClure1956,Raoux2015PRB}, and the proximity of a high-order VHS that should result in strong paramagnetism. The competition between these two
opposite responses together with the weak role of fourth band leads to a dia- to paramagnetic transition.

\subsection{The role of van Hove singularities}
\label{sec:susceptibility-van-Hove}
Let us discuss the role of van Hove singularities in T-graphene. For the ordinary VHS the orbital susceptibility exhibits paramagnetic peak \cite{Vignale1991PRL}. This can be understood using the standard Landau-Peierls formula for contribution of single band \cite{Raoux2015PRB}.
In T-graphene, at the doping level $\mu=\pm t_1$, one meets the three-band-touching points, at which two Dirac cones and middle band with flat lines
intersect.
In a single-layer graphene the presence of Dirac cones leads to singular diamagnetic contribution into orbital susceptibility $\chi\sim-\chi_0 \delta(\mu)$ at
zero temperature \cite{McClure1956}. In the gapped dice model, spectrum of which is similar to (\ref{spectrum:effective-linearH}), the paramagnetic contribution due to a flat band exceeds diamagnetic contribution from Dirac cones (see Ref.\cite{Piechon2015JPCM}).  In the case of T-graphene, the presence of middle band, which is
not flat anymore but contains flat lines with high-order VH singularities on it, leads to  strong paramagnetic contribution
competing with diamagnetic contribution from Dirac cones, thus resulting in sign change of the orbital susceptibility.

High-order Van Hove singularities  manifest themselves in many physical quantities as was reported in, e.g., Refs. \cite{Shtyk2017,Efremov2019,Ramires2012PRL,Classen2020,Lin2020arxiv,Sherkunov2018,Chichinadze2020,Isobe2018PRX,Wang2020arxiv,Lin2019highTc,Gonzalez2019}. In the present paper we focused on  the magnetic susceptibility of non-interacting electrons in square-octagon lattice. However, one should expect the manifestation of high-order VHS of T-graphene also in other physical quantities besides orbital susceptibility which is a subject for future studies.
We note that the accessibility of doping levels beyond the van Hove singularity was demonstrated in recent experiment for single-layer graphene \cite{Rosenzweig2020}.

\section{Conclusions }
\label{Conclusion}
In this paper we have studied the spectrum structure of tight-binding model for square-octagon lattice and  analyzed the emergence of Dirac cones and van Hove singularities of different type. Firstly, we found that the singularities in DOS, that correspond to the flat lines in spectrum of T-graphene, represent
VHS of high-order. Their benchmarks are large divergence exponent $\kappa=1/2$ (instead of logarithmic divergence for ordinary VHS) and
asymmetry of DOS near corresponding energy level. Such high-order saddle points in spectrum are intermediate between the ordinary saddle points and completely
flat bands. Also, using the L\"{o}wdin partitioning, we derived an effective second-order Hamiltonian that correctly captures dispersions of three bands near the high-order saddle point.

Secondly, we have studied the orbital susceptibility of electrons on square-octagon lattice. We have found that while for ordinary VHS there
are standard paramagnetic peaks predicted long ago by Vignale \cite{Vignale1991PRL}, the recently introduced high-order VHS \cite{Yuan2019Nature} manifest
themselves in a more complicated way. The tight-binding magnetic susceptibility exhibits several dia- to paramagnetic transitions when
a chemical potential runs the whole zone. 

Studying the orbital susceptibility at band-touching point ($\mu=t_1$) as a function of the tight-binding hoppings ratio $\alpha$, we found
a dia- to paramagnetic transition at $\alpha\approx0.94$. Its existence can be qualitatively understood due to competitions of contributions from
Dirac cones, which give strong diamagnetism,  and high-order VHS that result in strong paramagnetism. The effective low-energy pseudospin-1 Hamiltonian
near the $\Gamma$ point (\ref{eq:hamiltonian-Yamashita}) correctly describes paramagnetic contribution but  does not capture the dia- to paramagnetic transition. On the other hand, the effective Hamiltonian \eqref{eq:Yamashita-Lodwin-a-all}, which keeps second-order terms in a wavevector expansion,  correctly reproduces the dia- to paramagnetic transition at $\alpha=0.94$ given by the tight-binding Hamiltonian. 

The tight-binding parameter $\alpha$ can be varied due to in-plane deformations keeping $C_4$ symmetry, thus allowing to verify the dia- to  paramagnetic
transition in experiment. Though it is not probably  easy to fine-tune the hopping parameters experimentally, one can observe the different phases by analyzing different materials that are based on square-octagon lattice (see Refs. \cite{Gaikwad2017,Gaikwad2020, Majidi2015}). Also, the T-graphene model can be realized experimentally with cold fermionic atoms in an optical lattice, or in phononic crystals \cite{Choudhry2019PRB}. In these cases it could be possible to test directly  the sign change of the susceptibility as
a function of $\alpha$. In further studies of the T-graphene model it would be interesting to include impurities and interactions.

{\it Note Added in Proof.} In the recent publication
\cite{Guerci2021} the role of high-order VHS in the orbital magnetic susceptibility was studied for twisted
bilayer graphene. These studies complement the analysis in the present work.

\acknowledgments
We are grateful to E.V. Gorbar for useful remarks.
The work of V.P.G. is supported by the National Research Foundation of Ukraine grant "Topological phases of matter and excitations in Dirac materials, Josephson junctions and magnets". V.M.L. acknowledges collaboration within the Ukrainian-Israeli Scientific Research Program of the Ministry of Education and Science of
Ukraine (MESU) and the Ministry of Science and Technology of the state of Israel (MOST).

\appendix
\section{Flat lines in dispersion of middle bands and lattice symmetry}
\label{appendix:flat-lines}
In this Appendix we show that the flat lines in spectrum are related to the $C_{4}$ point symmetry group. Also we show, that every point of flat line represents a high-order saddle point. Firstly, one can check that setting $k_x =0$ (or $k_y =0$)
in spectral equation \eqref{eq:spectral_eq}, it can be factorized:

\begin{align}
(\epsilon -1) \left(-\left(4 \alpha ^2+1\right) \epsilon +4 \alpha ^2 \cos (a
k_y)+\epsilon ^3+\epsilon ^2-1\right)=0.
\end{align}
Here we used scaled energy parameter $\epsilon$, measured in units of $t_1$. Thus, we find the middle band dispersion $\epsilon=1$, which describes a flat line. The same property of spectral equation holds true for $k_{x}a=\pm \pi $ and $k_y a=\pm \pi$ lines, with $\epsilon=-1$.

The wavevector in tight-binding Hamiltonian \eqref{eq:tight-binding-Hamiltonian} is measured from $\Gamma$ point. Performing the rotation to the basis of $C_{4}$ symmetry group  via the unitary matrix given in Eq.\eqref{eq:C-4v-basis}, we find the transformed Hamiltonian:
\begin{widetext}
	\begin{align}
		U_{C_{4v}}^{\dagger} H U_{C_{4v}}=\frac{t_1}{2} \left(
		\begin{array}{cccc}
		-4 \alpha -\cos \left(a k_x\right)-\cos \left(a k_y\right) & i \sqrt{2} \sin \left(a
		k_x\right) & i \sqrt{2} \sin \left(a k_y\right) & -\cos \left(a k_x\right)+\cos
		\left(a k_y\right) \\
		-i \sqrt{2} \sin \left(a k_x\right) & 2 \cos \left(a k_x\right) & 0 & -i \sqrt{2} \sin
		\left(a k_x\right) \\
		-i \sqrt{2} \sin \left(a k_y\right) & 0 & 2 \cos \left(a k_y\right) & i \sqrt{2} \sin
		\left(a k_y\right) \\
		-\cos \left(a k_x\right)+\cos \left(a k_y\right) & i \sqrt{2} \sin \left(a k_x\right)
		& -i \sqrt{2} \sin \left(a k_y\right) & 4 \alpha -\cos \left(a k_x\right)-\cos
		\left(a k_y\right) \\
		\end{array}
		\right).
	\end{align}
\end{widetext}
It can be clearly seen that along flat line direction $k_x =0$ (and similarly for $k_y  = 0$), the Hamiltonian reduces to the matrix
\begin{widetext}
\begin{align}
U_{C_{4v}}^{\dagger} H U_{C_{4v}}(k_x=0, k_y)=
\frac{t_1}{2} \left(
	\begin{array}{cccc}
	-1-4 \alpha -\cos \left(a k_y\right) & 0 & i \sqrt{2} \sin \left(a k_y\right) & -1+\cos
	\left(a k_y\right) \\
	0 & 2 & 0 & 0 \\
	-i \sqrt{2} \sin \left(a k_y\right) & 0 & 2 \cos \left(a k_y\right) & i \sqrt{2} \sin
	\left(a k_y\right) \\
	-1+\cos \left(a k_y\right) & 0 & -i \sqrt{2} \sin \left(a k_y\right) & -1+4 \alpha -\cos
	\left(a k_y\right) \\
	\end{array}
	\right).
\end{align}
\end{widetext}
Thus, one can conclude that the presence of flat lines is protected not only by $C_{4}$ symmetry, but also by the geometry of tight-binding model.
As was noted in Ref.\cite{Yamashita2013PRB}, at the $\Gamma$ point the flat lines represent nearly flat band (two lines intersect at the angle
$\frac{\pi}{2}$). When the two hopping parameters are equal, $\alpha=1$, the corresponding linear low-energy model (\ref{eq:hamiltonian-Yamashita})
treats the middle band as completely flat and is similar to a pseudospin-1 model.
However, in the second order approximation (see Eq.\eqref{eq:Yamashita-Lodwin-a-all}) the middle band becomes dispersive. This fact distinguishes this
pseudospin-1 model from other models, such as Lieb \cite{Shen2010PRB}, Kagome \cite{Green2010PRB} or $\alpha-T_{3}$ \cite{Raoux2014,Gorbar2019PRB,Oriekhov2018LTP} models, where
the presence of exactly flat band is supported  by the lattice
geometry in tight-binding approximation.

Finally, expanding the spectral equation \eqref{eq:spectral_eq} near the flat line $k_x=0$ up to second order in $k_x a$, we find
\begin{align}
	&\delta ^4-4 \delta ^3+4\left(1- \alpha ^2\right) \delta ^2+2 \alpha ^2 \delta  \left((k_x a)^2-2 \cos (k_y a)+2\right)\nn
	&+2
	\alpha ^2 (k_x a)^2 (\cos (k_y a)-1)=0.
\end{align}
Here $\delta=1-\epsilon$ measures the deviation of energy from flat line value. In this equation we can omit the third and fourth order corrections ($\delta^3$
and $\delta^4$), and obtain simple quadratic equation. The solution, that corresponds to the flat line, has the following approximate behavior
\begin{align}
	\delta\approx \frac{k_x^2 a^2}{2}-\frac{k_x^4 a^4}{4 \alpha ^2 \left(\cos \left(k_y a\right)-1\right)}.
\end{align}
The determinant of Hessian matrix for such a solution is always zero. Thus we conclude, that every point on a flat line is a high-order saddle point.

\section{Green's function of tight-binding and L\"{o}wdin Hamiltonians}
\label{sec:green-func}
In this Appendix we calculate the Green function of the tight-binding Hamiltonian \eqref{eq:tight-binding-Hamiltonian}.
Standardly it is defined as
\be
G(\vec{k},\epsilon)=\frac{1}{t_1}\left(\epsilon-\frac{1}{t_1} H(\vec{k})\right)^{-1}
\ee
 for energy $\epsilon$ measured in units of $t_1$. Using the formula for adjoint matrix, we find the simple but long expression. For the clarity, we write the Green's function
 in block form:
\begin{align}
G(\vec{k},\epsilon)=\frac{1}{t_1 \det[\epsilon-\frac{1}{t_1}H(\vec{k})]}\left(\begin{array}{cc}
G_{11} & G_{12}\\
G_{12}^{\dagger} & G_{22}
\end{array}\right).
\end{align}
The corresponding blocks are given by the following expressions:
\begin{widetext}
	\begin{align}
	&G_{11}(\vec{k},\epsilon)=\left(
	\begin{array}{cc}
	\epsilon  \left(-2 \alpha ^2+\epsilon ^2-1\right)+2 \alpha ^2 \cos \left(k_y a\right) & \alpha  e^{-i k_y a} \left(-\epsilon
	+e^{i k_x a}\right) \left(-1+\epsilon e^{i k_y a}\right) \\
	\alpha  e^{-i k_x a} \left(-1+\epsilon  e^{i k_x a}\right) \left(-\epsilon +e^{i k_y a}\right) & \epsilon  \left(-2 \alpha
	^2+\epsilon^2-1\right)+2 \alpha ^2 \cos \left(k_x a\right) \\
	\end{array}
	\right)\\
	&G_{12}(\vec{k},\epsilon)=\left(
	\begin{array}{cc}
	2 \alpha ^2 \left(\epsilon -\cos \left(k_y a\right)\right)-\left(\epsilon^2-1\right) e^{i k_x a} & \alpha  \left(-\epsilon
	+e^{i k_x a}\right) \left(\epsilon -e^{i k_y a}\right) \\
	\alpha  \left(-\epsilon +e^{i k_x a}\right) \left(\epsilon -e^{i k_y a}\right) & 2 \alpha ^2 \left(\epsilon -\cos
	\left(k_x a\right)\right)-\left(\epsilon^2-1\right) e^{i k_y a} \\
	\end{array}
	\right)\\
	&G_{22}(\vec{k},\epsilon)=\left(
	\begin{array}{cc}
	\epsilon  \left(-2 \alpha ^2+\epsilon^2-1\right)+2 \alpha ^2 \cos \left(k_y a\right) & \alpha  e^{-i k_x a} \left(-1+\epsilon
	e^{i k_x a}\right) \left(-\epsilon+e^{i k_y a}\right) \\
	\alpha  e^{-i k_y a} \left(-\epsilon+e^{i k_x a}\right) \left(-1+\epsilon  e^{i k_y a}\right) & \epsilon  \left(-2 \alpha
	^2+\epsilon^2-1\right)+2 \alpha ^2 \cos \left(k_x a\right) \\
	\end{array}
	\right).
	\end{align}
\end{widetext}
These expressions are used to evaluate the traces for ``\textit{term 1}'' and ``\textit{term 2}'' (first and second terms in Eq.(\ref{X-matrix})):
 \begin{widetext}
 	\begin{align}\label{eq:term1-polynomial}
 		\tr[\text{term 1}]&=\frac{ a^4}{\det[\epsilon-\frac{1}{t_1}H(\vec{k})]^2}\bigg[4 \alpha ^2 \left(\left(\epsilon^2+1\right)
 		\cos (k_x a)-2 \epsilon\right)
 		\left(\left(\epsilon^2+1\right) \cos
 		(k_y a)-2 \epsilon \right)\bigg],\\
 		\label{eq:term2-polynomial}
 		\tr[\text{term 2}]&=\frac{16 \alpha ^2 a^4}{\det[\epsilon -\frac{1}{t_1}H(\vec{k})]^3}\bigg[ t_1^2 \alpha ^2 \left(\epsilon^2+2\right)^2+\epsilon
 		\left( \alpha ^2\epsilon \left(\epsilon^2+2\right) \cos (2k_x a)+\left((\epsilon^2-1)^2-4 \epsilon^2 \alpha ^2 \left(\epsilon
 		^2+2\right)\right) \cos (k_x a)\right)\bigg.\nn
 		&+\bigg. 2 \alpha^2 \epsilon^2 \cos (2 k_y a) (\epsilon\cos (k_x a)-1)^2+\cos (k_y a)\left(-2 \left(2 \alpha ^2+1\right) \epsilon^3-8 \alpha ^2
 		\epsilon -4 \alpha ^2\epsilon^3 \cos (2 k_xa)\right.\bigg.\nn
 		&+\bigg.\left.\left(4\alpha  \epsilon -\epsilon^2+1\right)\left(4 \alpha \epsilon +\epsilon^2-1\right) \cos (k_x a)+\epsilon^5+\epsilon \right)
 		-\epsilon^2\left(\epsilon^2-1\right)^2\bigg].
 	\end{align}
 \end{widetext}
For the second-order effective Hamiltonian \eqref{eq:Yamashita-Lodwin-a-all}, which is obtained with the help of L\"{o}wdin partitioning method, the Green's function is (we set $a=1$ to simplify the notation)
\begin{widetext}
	\begin{align}\label{eq:green-lodwin}
		&G=\frac{1}{t_1 \det\left[\epsilon-\frac{H_{SU(3)}^{(2)}(\vec{k})}{t_1}\right]}\times\\
		&\left(
		\begin{array}{ccc}
			\left[1-\frac{\vec{k^2}}{4}-2 \alpha +\epsilon \right] \left[\frac{(2 \alpha +1) k_y^2}{4 (\alpha +1)}+\epsilon
			-1\right]-\frac{k_y^2}{2} & -\frac{k_x k_y \left(\vec{k}^2+16 \alpha -4 \epsilon +4\right)}{16 (\alpha +1)} & -\frac{i k_x
				\left(2 (\epsilon -1)+\alpha  \left(k_y^2+2 \epsilon -2\right)\right)}{2 \sqrt{2} (\alpha +1)} \\
			-\frac{k_x k_y \left(\vec{k}^2+16 \alpha -4 \epsilon +4\right)}{16 (\alpha +1)} & \left[\frac{(2 \alpha +1) k_x^2}{4 (\alpha
				+1)}+\epsilon -1\right] \left[1-\frac{\vec{k}^2}{4}-2 \alpha +\epsilon\right]-\frac{k_x^2}{2} & \frac{i \left(2
				(\epsilon -1)+\alpha  \left(k_x^2+2 \epsilon -2\right)\right) k_y}{2 \sqrt{2} (\alpha +1)} \\
			\frac{i k_x \left(2 (\epsilon -1)+\alpha  \left(k_y^2+2 \epsilon -2\right)\right)}{2 \sqrt{2} (\alpha +1)} & -\frac{i \left(2
				(\epsilon -1)+\alpha  \left(k_x^2+2 \epsilon -2\right)\right) k_y}{2 \sqrt{2} (\alpha +1)} &(\epsilon
			-1)^2+\frac{(2 \alpha +1) \vec{k}^2 (\epsilon -1)+\alpha  k_x^2 k_y^2}{4 (\alpha +1)} \\
		\end{array}
		\right)\nonumber
	\end{align}
\end{widetext}
and the determinant is given by the following third-order polynomial:
\begin{widetext}
\begin{align}\label{eq:lodwin-det}
&\det\left[\epsilon-\frac{H_{SU(3)}^{(2)}(\vec{k})}{t_1}\right]=\epsilon ^3-\frac{\epsilon ^2 \left(\alpha  \left(8 \alpha -k^2+12\right)+\
4\right)}{4 (\alpha +1)}-\frac{\epsilon  \left(-32 (\alpha +1) (4\alpha -1)+\alpha  k^4 \cos (4 \phi )+(3 \alpha +2) k^4+16 \alpha  (2 \alpha +1) 
k^2\right)}{32 (\alpha +1)}\nn
&+\frac{-128\left(2 \alpha ^2+\alpha -1\right)-\alpha  k^6-4 (\alpha -2) (2 \alpha +1) k^4+32 \alpha  (4 \alpha +1) k^2+\alpha  k^4
\left(8 \alpha +k^2+4\right) \cos (4 \phi )}{128 (\alpha +1)}.
\end{align}
\end{widetext}

\bibliography{TgrapheneBib}

\begin{thebibliography}{57}%
\makeatletter
\providecommand \@ifxundefined [1]{%
 \@ifx{#1\undefined}
}%
\providecommand \@ifnum [1]{%
 \ifnum #1\expandafter \@firstoftwo
 \else \expandafter \@secondoftwo
 \fi
}%
\providecommand \@ifx [1]{%
 \ifx #1\expandafter \@firstoftwo
 \else \expandafter \@secondoftwo
 \fi
}%
\providecommand \natexlab [1]{#1}%
\providecommand \enquote  [1]{``#1''}%
\providecommand \bibnamefont  [1]{#1}%
\providecommand \bibfnamefont [1]{#1}%
\providecommand \citenamefont [1]{#1}%
\providecommand \href@noop [0]{\@secondoftwo}%
\providecommand \href [0]{\begingroup \@sanitize@url \@href}%
\providecommand \@href[1]{\@@startlink{#1}\@@href}%
\providecommand \@@href[1]{\endgroup#1\@@endlink}%
\providecommand \@sanitize@url [0]{\catcode `\\12\catcode `\$12\catcode
  `\&12\catcode `\#12\catcode `\^12\catcode `\_12\catcode `\%12\relax}%
\providecommand \@@startlink[1]{}%
\providecommand \@@endlink[0]{}%
\providecommand \url  [0]{\begingroup\@sanitize@url \@url }%
\providecommand \@url [1]{\endgroup\@href {#1}{\urlprefix }}%
\providecommand \urlprefix  [0]{URL }%
\providecommand \Eprint [0]{\href }%
\providecommand \doibase [0]{https://doi.org/}%
\providecommand \selectlanguage [0]{\@gobble}%
\providecommand \bibinfo  [0]{\@secondoftwo}%
\providecommand \bibfield  [0]{\@secondoftwo}%
\providecommand \translation [1]{[#1]}%
\providecommand \BibitemOpen [0]{}%
\providecommand \bibitemStop [0]{}%
\providecommand \bibitemNoStop [0]{.\EOS\space}%
\providecommand \EOS [0]{\spacefactor3000\relax}%
\providecommand \BibitemShut  [1]{\csname bibitem#1\endcsname}%
\let\auto@bib@innerbib\@empty
\bibitem [{\citenamefont {Liu}\ \emph {et~al.}(2012)\citenamefont {Liu},
  \citenamefont {Wang}, \citenamefont {Huang}, \citenamefont {Guo},\ and\
  \citenamefont {Chen}}]{Liu2012PRL}%
  \BibitemOpen
  \bibfield  {author} {\bibinfo {author} {\bibfnamefont {Y.}~\bibnamefont
  {Liu}}, \bibinfo {author} {\bibfnamefont {G.}~\bibnamefont {Wang}}, \bibinfo
  {author} {\bibfnamefont {Q.}~\bibnamefont {Huang}}, \bibinfo {author}
  {\bibfnamefont {L.}~\bibnamefont {Guo}},\ and\ \bibinfo {author}
  {\bibfnamefont {X.}~\bibnamefont {Chen}},\ }\bibfield  {title} {\bibinfo
  {title} {{S}tructural and {E}lectronic {P}roperties of ${T}$ {G}raphene: {A}
  {T}wo-{D}imensional {C}arbon {A}llotrope with {T}etrarings},\ }\href
  {https://doi.org/10.1103/PhysRevLett.108.225505} {\bibfield  {journal}
  {\bibinfo  {journal} {Physical Review Letters}\ }\textbf {\bibinfo {volume}
  {108}},\ \bibinfo {pages} {225505} (\bibinfo {year} {2012})}\BibitemShut
  {NoStop}%
\bibitem [{\citenamefont {Terrones}\ \emph {et~al.}(2000)\citenamefont
  {Terrones}, \citenamefont {Terrones}, \citenamefont {Hern\'andez},
  \citenamefont {Grobert}, \citenamefont {Charlier},\ and\ \citenamefont
  {Ajayan}}]{Terrones2000}%
  \BibitemOpen
  \bibfield  {author} {\bibinfo {author} {\bibfnamefont {H.}~\bibnamefont
  {Terrones}}, \bibinfo {author} {\bibfnamefont {M.}~\bibnamefont {Terrones}},
  \bibinfo {author} {\bibfnamefont {E.}~\bibnamefont {Hern\'andez}}, \bibinfo
  {author} {\bibfnamefont {N.}~\bibnamefont {Grobert}}, \bibinfo {author}
  {\bibfnamefont {J.-C.}\ \bibnamefont {Charlier}},\ and\ \bibinfo {author}
  {\bibfnamefont {P.~M.}\ \bibnamefont {Ajayan}},\ }\bibfield  {title}
  {\bibinfo {title} {New metallic allotropes of planar and tubular carbon},\
  }\href {https://doi.org/10.1103/PhysRevLett.84.1716} {\bibfield  {journal}
  {\bibinfo  {journal} {Phys. Rev. Lett.}\ }\textbf {\bibinfo {volume} {84}},\
  \bibinfo {pages} {1716} (\bibinfo {year} {2000})}\BibitemShut {NoStop}%
\bibitem [{\citenamefont {Enyashin}\ and\ \citenamefont
  {Ivanovskii}(2011)}]{Enyashin2011}%
  \BibitemOpen
  \bibfield  {author} {\bibinfo {author} {\bibfnamefont {A.~N.}\ \bibnamefont
  {Enyashin}}\ and\ \bibinfo {author} {\bibfnamefont {A.~L.}\ \bibnamefont
  {Ivanovskii}},\ }\bibfield  {title} {\bibinfo {title} {Graphene allotropes},\
  }\href {https://doi.org/10.1002/pssb.201046583} {\bibfield  {journal}
  {\bibinfo  {journal} {Physica Status Solidi (b)}\ }\textbf {\bibinfo {volume}
  {248}},\ \bibinfo {pages} {1879} (\bibinfo {year} {2011})}\BibitemShut
  {NoStop}%
\bibitem [{\citenamefont {Kim}\ \emph {et~al.}(2013)\citenamefont {Kim},
  \citenamefont {Jo},\ and\ \citenamefont {Sim}}]{Kim2013}%
  \BibitemOpen
  \bibfield  {author} {\bibinfo {author} {\bibfnamefont {B.~G.}\ \bibnamefont
  {Kim}}, \bibinfo {author} {\bibfnamefont {J.~Y.}\ \bibnamefont {Jo}},\ and\
  \bibinfo {author} {\bibfnamefont {H.~S.}\ \bibnamefont {Sim}},\ }\bibfield
  {title} {\bibinfo {title} {Comment on ``{S}tructural and {E}lectronic
  {P}roperties of ${T}$ {G}raphene: {A} {T}wo-{D}imensional {C}arbon
  {A}llotrope with {T}etrarings''},\ }\href
  {https://doi.org/10.1103/PhysRevLett.110.029601} {\bibfield  {journal}
  {\bibinfo  {journal} {Phys. Rev. Lett.}\ }\textbf {\bibinfo {volume} {110}},\
  \bibinfo {pages} {029601} (\bibinfo {year} {2013})}\BibitemShut {NoStop}%
\bibitem [{\citenamefont {Qinyan~Gu}(2019)}]{QinyanGu:97401}%
  \BibitemOpen
  \bibfield  {author} {\bibinfo {author} {\bibfnamefont {J.~S.}\ \bibnamefont
  {Qinyan~Gu}, \bibfnamefont {Dingyu~Xing}},\ }\bibfield  {title} {\bibinfo
  {title} {Superconducting single-layer {T}-graphene and novel synthesis
  routes},\ }\href {https://doi.org/10.1088/0256-307X/36/9/097401} {\bibfield
  {journal} {\bibinfo  {journal} {Chinese Physics Letters}\ }\textbf {\bibinfo
  {volume} {36}},\ \bibinfo {eid} {097401} (\bibinfo {year}
  {2019})}\BibitemShut {NoStop}%
\bibitem [{\citenamefont {Sheng}\ \emph {et~al.}(2012)\citenamefont {Sheng},
  \citenamefont {Cui}, \citenamefont {Ye}, \citenamefont {Yan}, \citenamefont
  {Zheng},\ and\ \citenamefont {Su}}]{Sheng2012}%
  \BibitemOpen
  \bibfield  {author} {\bibinfo {author} {\bibfnamefont {X.-L.}\ \bibnamefont
  {Sheng}}, \bibinfo {author} {\bibfnamefont {H.-J.}\ \bibnamefont {Cui}},
  \bibinfo {author} {\bibfnamefont {F.}~\bibnamefont {Ye}}, \bibinfo {author}
  {\bibfnamefont {Q.-B.}\ \bibnamefont {Yan}}, \bibinfo {author} {\bibfnamefont
  {Q.-R.}\ \bibnamefont {Zheng}},\ and\ \bibinfo {author} {\bibfnamefont
  {G.}~\bibnamefont {Su}},\ }\bibfield  {title} {\bibinfo {title} {Octagraphene
  as a versatile carbon atomic sheet for novel nanotubes, unconventional
  fullerenes, and hydrogen storage},\ }\href
  {https://doi.org/10.1063/1.4757410} {\bibfield  {journal} {\bibinfo
  {journal} {Journal of Applied Physics}\ }\textbf {\bibinfo {volume} {112}},\
  \bibinfo {pages} {074315} (\bibinfo {year} {2012})}\BibitemShut {NoStop}%
\bibitem [{\citenamefont {Bao}\ \emph {et~al.}(2014)\citenamefont {Bao},
  \citenamefont {Tao}, \citenamefont {Liu}, \citenamefont {Zhang},\ and\
  \citenamefont {Liu}}]{Bao2014Nature}%
  \BibitemOpen
  \bibfield  {author} {\bibinfo {author} {\bibfnamefont {A.}~\bibnamefont
  {Bao}}, \bibinfo {author} {\bibfnamefont {H.-S.}\ \bibnamefont {Tao}},
  \bibinfo {author} {\bibfnamefont {H.-D.}\ \bibnamefont {Liu}}, \bibinfo
  {author} {\bibfnamefont {X.}~\bibnamefont {Zhang}},\ and\ \bibinfo {author}
  {\bibfnamefont {W.-M.}\ \bibnamefont {Liu}},\ }\bibfield  {title} {\bibinfo
  {title} {Quantum magnetic phase transition in square-octagon lattice},\
  }\href {https://doi.org/10.1038/srep06918} {\bibfield  {journal} {\bibinfo
  {journal} {Scientific Reports}\ }\textbf {\bibinfo {volume} {4}},\ \bibinfo
  {pages} {6918} (\bibinfo {year} {2014})}\BibitemShut {NoStop}%
\bibitem [{\citenamefont {Yamashita}\ \emph {et~al.}(2013)\citenamefont
  {Yamashita}, \citenamefont {Tomura}, \citenamefont {Yanagi},\ and\
  \citenamefont {Ueda}}]{Yamashita2013PRB}%
  \BibitemOpen
  \bibfield  {author} {\bibinfo {author} {\bibfnamefont {Y.}~\bibnamefont
  {Yamashita}}, \bibinfo {author} {\bibfnamefont {M.}~\bibnamefont {Tomura}},
  \bibinfo {author} {\bibfnamefont {Y.}~\bibnamefont {Yanagi}},\ and\ \bibinfo
  {author} {\bibfnamefont {K.}~\bibnamefont {Ueda}},\ }\bibfield  {title}
  {\bibinfo {title} {{S}{U}(3) {D}irac electrons in the $\frac{1}{5}$-depleted
  square-lattice {H}ubbard model at $\frac{1}{4}$ filling},\ }\href
  {https://doi.org/10.1103/PhysRevB.88.195104} {\bibfield  {journal} {\bibinfo
  {journal} {Physical Review B}\ }\textbf {\bibinfo {volume} {88}},\ \bibinfo
  {pages} {195104} (\bibinfo {year} {2013})}\BibitemShut {NoStop}%
\bibitem [{\citenamefont {Li}\ \emph {et~al.}(2014)\citenamefont {Li},
  \citenamefont {Guo}, \citenamefont {Zhang},\ and\ \citenamefont
  {Zhang}}]{Li2014}%
  \BibitemOpen
  \bibfield  {author} {\bibinfo {author} {\bibfnamefont {W.}~\bibnamefont
  {Li}}, \bibinfo {author} {\bibfnamefont {M.}~\bibnamefont {Guo}}, \bibinfo
  {author} {\bibfnamefont {G.}~\bibnamefont {Zhang}},\ and\ \bibinfo {author}
  {\bibfnamefont {Y.-W.}\ \bibnamefont {Zhang}},\ }\bibfield  {title} {\bibinfo
  {title} {Gapless {M}o{S}$_{2}$ allotrope possessing both massless {D}irac and
  heavy fermions},\ }\href {https://doi.org/10.1103/PhysRevB.89.205402}
  {\bibfield  {journal} {\bibinfo  {journal} {Phys. Rev. B}\ }\textbf {\bibinfo
  {volume} {89}},\ \bibinfo {pages} {205402} (\bibinfo {year}
  {2014})}\BibitemShut {NoStop}%
\bibitem [{\citenamefont {Pomata}\ and\ \citenamefont
  {Wei}(2019)}]{Pomata2019}%
  \BibitemOpen
  \bibfield  {author} {\bibinfo {author} {\bibfnamefont {N.}~\bibnamefont
  {Pomata}}\ and\ \bibinfo {author} {\bibfnamefont {T.-C.}\ \bibnamefont
  {Wei}},\ }\bibfield  {title} {\bibinfo {title} {Demonstrating the
  {A}{K}{L}{T} spectral gap on 2{D} degree-3 lattices},\ }\href@noop {}
  {\bibfield  {journal} {\bibinfo  {journal} {arxiv:1911.01410}\ } (\bibinfo
  {year} {2019})},\ \Eprint {https://arxiv.org/abs/1911.01410v2} {1911.01410v2}
  \BibitemShut {NoStop}%
\bibitem [{\citenamefont {Sun}\ \emph {et~al.}(2015)\citenamefont {Sun},
  \citenamefont {Felser},\ and\ \citenamefont {Yan}}]{Sun2015}%
  \BibitemOpen
  \bibfield  {author} {\bibinfo {author} {\bibfnamefont {Y.}~\bibnamefont
  {Sun}}, \bibinfo {author} {\bibfnamefont {C.}~\bibnamefont {Felser}},\ and\
  \bibinfo {author} {\bibfnamefont {B.}~\bibnamefont {Yan}},\ }\bibfield
  {title} {\bibinfo {title} {Graphene-like {D}irac states and quantum spin
  {H}all insulators in square-octagonal ${M}{X}_{2}$ (${M}=\mathrm{Mo},\,
  \mathrm{W}$; ${X}=\mathrm{S},\, \mathrm{Se},\, \mathrm{Te}$) isomers},\
  }\href {https://doi.org/10.1103/physrevb.92.165421} {\bibfield  {journal}
  {\bibinfo  {journal} {Physical Review B}\ }\textbf {\bibinfo {volume} {92}},\
  \bibinfo {pages} {165421} (\bibinfo {year} {2015})}\BibitemShut {NoStop}%
\bibitem [{\citenamefont {Umemoto}\ \emph {et~al.}(2010)\citenamefont
  {Umemoto}, \citenamefont {Wentzcovitch}, \citenamefont {Saito},\ and\
  \citenamefont {Miyake}}]{Umemoto2010}%
  \BibitemOpen
  \bibfield  {author} {\bibinfo {author} {\bibfnamefont {K.}~\bibnamefont
  {Umemoto}}, \bibinfo {author} {\bibfnamefont {R.~M.}\ \bibnamefont
  {Wentzcovitch}}, \bibinfo {author} {\bibfnamefont {S.}~\bibnamefont
  {Saito}},\ and\ \bibinfo {author} {\bibfnamefont {T.}~\bibnamefont
  {Miyake}},\ }\bibfield  {title} {\bibinfo {title} {{B}ody-{C}entered
  {T}etragonal $\mathbf{C}_{4}$: {A} {V}iable $s{p}^{3}$ {C}arbon
  {A}llotrope},\ }\href {https://doi.org/10.1103/PhysRevLett.104.125504}
  {\bibfield  {journal} {\bibinfo  {journal} {Physical Review Letters}\
  }\textbf {\bibinfo {volume} {104}},\ \bibinfo {pages} {125504} (\bibinfo
  {year} {2010})}\BibitemShut {NoStop}%
\bibitem [{\citenamefont {Majidi}(2015)}]{Majidi2015}%
  \BibitemOpen
  \bibfield  {author} {\bibinfo {author} {\bibfnamefont {R.}~\bibnamefont
  {Majidi}},\ }\bibfield  {title} {\bibinfo {title} {Electronic properties of
  {T} graphene-like {C}{\textendash}{BN} sheets: {A} density functional theory
  study},\ }\href {https://doi.org/10.1016/j.physe.2015.07.029} {\bibfield
  {journal} {\bibinfo  {journal} {Phys. E: Low-dim. Sys. and Nanostr.}\
  }\textbf {\bibinfo {volume} {74}},\ \bibinfo {pages} {371} (\bibinfo {year}
  {2015})}\BibitemShut {NoStop}%
\bibitem [{\citenamefont {Majidi}(2017)}]{Majidi2017}%
  \BibitemOpen
  \bibfield  {author} {\bibinfo {author} {\bibfnamefont {R.}~\bibnamefont
  {Majidi}},\ }\bibfield  {title} {\bibinfo {title} {Density functional theory
  study on structural and mechanical properties of graphene, {T}-graphene, and
  {R}-graphyne},\ }\href {https://doi.org/10.1007/s00214-017-2148-1} {\bibfield
   {journal} {\bibinfo  {journal} {Theoretical Chemistry Accounts}\ }\textbf
  {\bibinfo {volume} {136}},\ \bibinfo {pages} {109} (\bibinfo {year}
  {2017})}\BibitemShut {NoStop}%
\bibitem [{\citenamefont {Yin}\ \emph {et~al.}(2013)\citenamefont {Yin},
  \citenamefont {Xie}, \citenamefont {Liu}, \citenamefont {Wang}, \citenamefont
  {Wei}, \citenamefont {Lau}, \citenamefont {Zhong},\ and\ \citenamefont
  {Chen}}]{Yin2013}%
  \BibitemOpen
  \bibfield  {author} {\bibinfo {author} {\bibfnamefont {W.-J.}\ \bibnamefont
  {Yin}}, \bibinfo {author} {\bibfnamefont {Y.-E.}\ \bibnamefont {Xie}},
  \bibinfo {author} {\bibfnamefont {L.-M.}\ \bibnamefont {Liu}}, \bibinfo
  {author} {\bibfnamefont {R.-Z.}\ \bibnamefont {Wang}}, \bibinfo {author}
  {\bibfnamefont {X.-L.}\ \bibnamefont {Wei}}, \bibinfo {author} {\bibfnamefont
  {L.}~\bibnamefont {Lau}}, \bibinfo {author} {\bibfnamefont {J.-X.}\
  \bibnamefont {Zhong}},\ and\ \bibinfo {author} {\bibfnamefont {Y.-P.}\
  \bibnamefont {Chen}},\ }\bibfield  {title} {\bibinfo {title} {R-graphyne: a
  new two-dimensional carbon allotrope with versatile {D}irac-like point in
  nanoribbons},\ }\href {https://doi.org/10.1039/c3ta00097d} {\bibfield
  {journal} {\bibinfo  {journal} {Journal of Materials Chemistry A}\ }\textbf
  {\bibinfo {volume} {1}},\ \bibinfo {pages} {5341} (\bibinfo {year}
  {2013})}\BibitemShut {NoStop}%
\bibitem [{\citenamefont {Podlivaev}\ and\ \citenamefont
  {Openov}(2013)}]{Podlivaev2013}%
  \BibitemOpen
  \bibfield  {author} {\bibinfo {author} {\bibfnamefont {A.~I.}\ \bibnamefont
  {Podlivaev}}\ and\ \bibinfo {author} {\bibfnamefont {L.~A.}\ \bibnamefont
  {Openov}},\ }\bibfield  {title} {\bibinfo {title} {Kinetic stability of
  octagraphene},\ }\href {https://doi.org/10.1134/s1063783413120299} {\bibfield
   {journal} {\bibinfo  {journal} {Physics of the Solid State}\ }\textbf
  {\bibinfo {volume} {55}},\ \bibinfo {pages} {2592} (\bibinfo {year}
  {2013})}\BibitemShut {NoStop}%
\bibitem [{\citenamefont {Gaikwad}\ \emph {et~al.}(2017)\citenamefont
  {Gaikwad}, \citenamefont {Pujari}, \citenamefont {Chakroborty},\ and\
  \citenamefont {Kshirsagar}}]{Gaikwad2017}%
  \BibitemOpen
  \bibfield  {author} {\bibinfo {author} {\bibfnamefont {P.~V.}\ \bibnamefont
  {Gaikwad}}, \bibinfo {author} {\bibfnamefont {P.~K.}\ \bibnamefont {Pujari}},
  \bibinfo {author} {\bibfnamefont {S.}~\bibnamefont {Chakroborty}},\ and\
  \bibinfo {author} {\bibfnamefont {A.}~\bibnamefont {Kshirsagar}},\ }\bibfield
   {title} {\bibinfo {title} {Cluster assembly route to a novel octagonal
  two-dimensional {ZnO} monolayer},\ }\href
  {https://doi.org/10.1088/1361-648x/aa787e} {\bibfield  {journal} {\bibinfo
  {journal} {Journal of Physics: Condensed Matter}\ }\textbf {\bibinfo {volume}
  {29}},\ \bibinfo {pages} {335501} (\bibinfo {year} {2017})}\BibitemShut
  {NoStop}%
\bibitem [{\citenamefont {Gaikwad}\ and\ \citenamefont
  {Kshirsagar}()}]{Gaikwad2020}%
  \BibitemOpen
  \bibfield  {author} {\bibinfo {author} {\bibfnamefont {P.~V.}\ \bibnamefont
  {Gaikwad}}\ and\ \bibinfo {author} {\bibfnamefont {A.}~\bibnamefont
  {Kshirsagar}},\ }\bibfield  {title} {\bibinfo {title} {Octagonal family of
  monolayers, bulk and nanotubes},\ }\href {https://arxiv.org/abs/2003.00158}
  {\bibinfo  {journal} {arxiv:2003.00158}\ }\BibitemShut {NoStop}%
\bibitem [{\citenamefont {Yuan}\ \emph {et~al.}(2019)\citenamefont {Yuan},
  \citenamefont {Isobe},\ and\ \citenamefont {Fu}}]{Yuan2019Nature}%
  \BibitemOpen
\bibfield  {journal} {  }\bibfield  {author} {\bibinfo {author} {\bibfnamefont
  {N.~F.~Q.}\ \bibnamefont {Yuan}}, \bibinfo {author} {\bibfnamefont
  {H.}~\bibnamefont {Isobe}},\ and\ \bibinfo {author} {\bibfnamefont
  {L.}~\bibnamefont {Fu}},\ }\bibfield  {title} {\bibinfo {title} {Magic of
  high-order van {H}ove singularity},\ }\href
  {https://doi.org/10.1038/s41467-019-13670-9} {\bibfield  {journal} {\bibinfo
  {journal} {Nature Communications}\ }\textbf {\bibinfo {volume} {10}},\
  \bibinfo {pages} {5769} (\bibinfo {year} {2019})}\BibitemShut {NoStop}%
\bibitem [{\citenamefont {Vignale}(1991)}]{Vignale1991PRL}%
  \BibitemOpen
  \bibfield  {author} {\bibinfo {author} {\bibfnamefont {G.}~\bibnamefont
  {Vignale}},\ }\bibfield  {title} {\bibinfo {title} {Orbital paramagnetism of
  electrons in a two-dimensional lattice},\ }\href
  {https://doi.org/10.1103/physrevlett.67.358} {\bibfield  {journal} {\bibinfo
  {journal} {Physical Review Letters}\ }\textbf {\bibinfo {volume} {67}},\
  \bibinfo {pages} {358} (\bibinfo {year} {1991})}\BibitemShut {NoStop}%
\bibitem [{\citenamefont {Nandkishore}\ \emph {et~al.}(2012)\citenamefont
  {Nandkishore}, \citenamefont {Levitov},\ and\ \citenamefont
  {Chubukov}}]{Nandkishore2012}%
  \BibitemOpen
  \bibfield  {author} {\bibinfo {author} {\bibfnamefont {R.}~\bibnamefont
  {Nandkishore}}, \bibinfo {author} {\bibfnamefont {L.~S.}\ \bibnamefont
  {Levitov}},\ and\ \bibinfo {author} {\bibfnamefont {A.~V.}\ \bibnamefont
  {Chubukov}},\ }\bibfield  {title} {\bibinfo {title} {Chiral superconductivity
  from repulsive interactions in doped graphene},\ }\href
  {https://doi.org/10.1038/nphys2208} {\bibfield  {journal} {\bibinfo
  {journal} {Nature Physics}\ }\textbf {\bibinfo {volume} {8}},\ \bibinfo
  {pages} {158} (\bibinfo {year} {2012})}\BibitemShut {NoStop}%
\bibitem [{\citenamefont {Isobe}\ and\ \citenamefont {Fu}(2019)}]{Isobe2019}%
  \BibitemOpen
  \bibfield  {author} {\bibinfo {author} {\bibfnamefont {H.}~\bibnamefont
  {Isobe}}\ and\ \bibinfo {author} {\bibfnamefont {L.}~\bibnamefont {Fu}},\
  }\bibfield  {title} {\bibinfo {title} {Supermetal},\ }\href
  {https://doi.org/10.1103/physrevresearch.1.033206} {\bibfield  {journal}
  {\bibinfo  {journal} {Physical Review Research}\ }\textbf {\bibinfo {volume}
  {1}},\ \bibinfo {pages} {033206} (\bibinfo {year} {2019})}\BibitemShut
  {NoStop}%
\bibitem [{\citenamefont {Bradlyn}\ \emph {et~al.}(2016)\citenamefont
  {Bradlyn}, \citenamefont {Cano}, \citenamefont {Wang}, \citenamefont
  {Vergniory}, \citenamefont {Felser}, \citenamefont {Cava},\ and\
  \citenamefont {Bernevig}}]{Bradlyn2016}%
  \BibitemOpen
  \bibfield  {author} {\bibinfo {author} {\bibfnamefont {B.}~\bibnamefont
  {Bradlyn}}, \bibinfo {author} {\bibfnamefont {J.}~\bibnamefont {Cano}},
  \bibinfo {author} {\bibfnamefont {Z.}~\bibnamefont {Wang}}, \bibinfo {author}
  {\bibfnamefont {M.~G.}\ \bibnamefont {Vergniory}}, \bibinfo {author}
  {\bibfnamefont {C.}~\bibnamefont {Felser}}, \bibinfo {author} {\bibfnamefont
  {R.~J.}\ \bibnamefont {Cava}},\ and\ \bibinfo {author} {\bibfnamefont
  {B.~A.}\ \bibnamefont {Bernevig}},\ }\bibfield  {title} {\bibinfo {title}
  {{B}eyond {D}irac and {W}eyl fermions: {U}nconventional quasiparticles in
  conventional crystals},\ }\href {https://doi.org/10.1126/science.aaf5037}
  {\bibfield  {journal} {\bibinfo  {journal} {Science}\ }\textbf {\bibinfo
  {volume} {353}},\ \bibinfo {pages} {aaf5037} (\bibinfo {year}
  {2016})}\BibitemShut {NoStop}%
\bibitem [{\citenamefont {Shtyk}\ \emph {et~al.}(2017)\citenamefont {Shtyk},
  \citenamefont {Goldstein},\ and\ \citenamefont {Chamon}}]{Shtyk2017}%
  \BibitemOpen
  \bibfield  {author} {\bibinfo {author} {\bibfnamefont {A.}~\bibnamefont
  {Shtyk}}, \bibinfo {author} {\bibfnamefont {G.}~\bibnamefont {Goldstein}},\
  and\ \bibinfo {author} {\bibfnamefont {C.}~\bibnamefont {Chamon}},\
  }\bibfield  {title} {\bibinfo {title} {Electrons at the monkey saddle: {A}
  multicritical {L}ifshitz point},\ }\href
  {https://doi.org/10.1103/physrevb.95.035137} {\bibfield  {journal} {\bibinfo
  {journal} {Physical Review B}\ }\textbf {\bibinfo {volume} {95}},\ \bibinfo
  {pages} {035137} (\bibinfo {year} {2017})}\BibitemShut {NoStop}%
\bibitem [{\citenamefont {Efremov}\ \emph {et~al.}(2019)\citenamefont
  {Efremov}, \citenamefont {Shtyk}, \citenamefont {Rost}, \citenamefont
  {Chamon}, \citenamefont {Mackenzie},\ and\ \citenamefont
  {Betouras}}]{Efremov2019}%
  \BibitemOpen
  \bibfield  {author} {\bibinfo {author} {\bibfnamefont {D.~V.}\ \bibnamefont
  {Efremov}}, \bibinfo {author} {\bibfnamefont {A.}~\bibnamefont {Shtyk}},
  \bibinfo {author} {\bibfnamefont {A.~W.}\ \bibnamefont {Rost}}, \bibinfo
  {author} {\bibfnamefont {C.}~\bibnamefont {Chamon}}, \bibinfo {author}
  {\bibfnamefont {A.~P.}\ \bibnamefont {Mackenzie}},\ and\ \bibinfo {author}
  {\bibfnamefont {J.~J.}\ \bibnamefont {Betouras}},\ }\bibfield  {title}
  {\bibinfo {title} {{M}ulticritical {F}ermi {S}urface {T}opological
  {T}ransitions},\ }\href {https://doi.org/10.1103/physrevlett.123.207202}
  {\bibfield  {journal} {\bibinfo  {journal} {Physical Review Letters}\
  }\textbf {\bibinfo {volume} {123}},\ \bibinfo {pages} {207202} (\bibinfo
  {year} {2019})}\BibitemShut {NoStop}%
\bibitem [{\citenamefont {Ramires}\ \emph {et~al.}(2012)\citenamefont
  {Ramires}, \citenamefont {Coleman}, \citenamefont {Nevidomskyy},\ and\
  \citenamefont {Tsvelik}}]{Ramires2012PRL}%
  \BibitemOpen
  \bibfield  {author} {\bibinfo {author} {\bibfnamefont {A.}~\bibnamefont
  {Ramires}}, \bibinfo {author} {\bibfnamefont {P.}~\bibnamefont {Coleman}},
  \bibinfo {author} {\bibfnamefont {A.~H.}\ \bibnamefont {Nevidomskyy}},\ and\
  \bibinfo {author} {\bibfnamefont {A.~M.}\ \bibnamefont {Tsvelik}},\
  }\bibfield  {title} {\bibinfo {title} {$\beta$-{YbAlB}4: A critical nodal
  metal},\ }\href {https://doi.org/10.1103/physrevlett.109.176404} {\bibfield
  {journal} {\bibinfo  {journal} {Physical Review Letters}\ }\textbf {\bibinfo
  {volume} {109}},\ \bibinfo {pages} {176404} (\bibinfo {year}
  {2012})}\BibitemShut {NoStop}%
\bibitem [{\citenamefont {Classen}\ \emph {et~al.}(2020)\citenamefont
  {Classen}, \citenamefont {Chubukov}, \citenamefont {Honerkamp},\ and\
  \citenamefont {Scherer}}]{Classen2020}%
  \BibitemOpen
  \bibfield  {author} {\bibinfo {author} {\bibfnamefont {L.}~\bibnamefont
  {Classen}}, \bibinfo {author} {\bibfnamefont {A.~V.}\ \bibnamefont
  {Chubukov}}, \bibinfo {author} {\bibfnamefont {C.}~\bibnamefont
  {Honerkamp}},\ and\ \bibinfo {author} {\bibfnamefont {M.~M.}\ \bibnamefont
  {Scherer}},\ }\bibfield  {title} {\bibinfo {title} {Competing orders at
  higher-order van hove points},\ }\href
  {https://doi.org/10.1103/PhysRevB.102.125141} {\bibfield  {journal} {\bibinfo
   {journal} {Phys. Rev. B}\ }\textbf {\bibinfo {volume} {102}},\ \bibinfo
  {pages} {125141} (\bibinfo {year} {2020})}\BibitemShut {NoStop}%
\bibitem [{\citenamefont {Lin}\ and\ \citenamefont
  {Nandkishore}()}]{Lin2020arxiv}%
  \BibitemOpen
  \bibfield  {author} {\bibinfo {author} {\bibfnamefont {Y.-P.}\ \bibnamefont
  {Lin}}\ and\ \bibinfo {author} {\bibfnamefont {R.~M.}\ \bibnamefont
  {Nandkishore}},\ }\bibfield  {title} {\bibinfo {title} {Parquet
  renormalization group analysis of weak-coupling instabilities with multiple
  high-order {V}an {H}ove points inside the {B}rillouin zone},\ }\href@noop {}
  {\ }\Eprint {https://arxiv.org/abs/2008.05485v1} {2008.05485v1} \BibitemShut
  {NoStop}%
\bibitem [{\citenamefont {Sherkunov}\ and\ \citenamefont
  {Betouras}(2018)}]{Sherkunov2018}%
  \BibitemOpen
  \bibfield  {author} {\bibinfo {author} {\bibfnamefont {Y.}~\bibnamefont
  {Sherkunov}}\ and\ \bibinfo {author} {\bibfnamefont {J.~J.}\ \bibnamefont
  {Betouras}},\ }\bibfield  {title} {\bibinfo {title} {Electronic phases in
  twisted bilayer graphene at magic angles as a result of van hove
  singularities and interactions},\ }\href
  {https://doi.org/10.1103/PhysRevB.98.205151} {\bibfield  {journal} {\bibinfo
  {journal} {Phys. Rev. B}\ }\textbf {\bibinfo {volume} {98}},\ \bibinfo
  {pages} {205151} (\bibinfo {year} {2018})}\BibitemShut {NoStop}%
\bibitem [{\citenamefont {Chichinadze}\ \emph {et~al.}(2020)\citenamefont
  {Chichinadze}, \citenamefont {Classen},\ and\ \citenamefont
  {Chubukov}}]{Chichinadze2020}%
  \BibitemOpen
  \bibfield  {author} {\bibinfo {author} {\bibfnamefont {D.~V.}\ \bibnamefont
  {Chichinadze}}, \bibinfo {author} {\bibfnamefont {L.}~\bibnamefont
  {Classen}},\ and\ \bibinfo {author} {\bibfnamefont {A.~V.}\ \bibnamefont
  {Chubukov}},\ }\bibfield  {title} {\bibinfo {title} {Valley magnetism,
  nematicity, and density wave orders in twisted bilayer graphene},\ }\href
  {https://doi.org/10.1103/PhysRevB.102.125120} {\bibfield  {journal} {\bibinfo
   {journal} {Phys. Rev. B}\ }\textbf {\bibinfo {volume} {102}},\ \bibinfo
  {pages} {125120} (\bibinfo {year} {2020})}\BibitemShut {NoStop}%
\bibitem [{\citenamefont {Isobe}\ \emph {et~al.}(2018)\citenamefont {Isobe},
  \citenamefont {Yuan},\ and\ \citenamefont {Fu}}]{Isobe2018PRX}%
  \BibitemOpen
  \bibfield  {author} {\bibinfo {author} {\bibfnamefont {H.}~\bibnamefont
  {Isobe}}, \bibinfo {author} {\bibfnamefont {N.~F.~Q.}\ \bibnamefont {Yuan}},\
  and\ \bibinfo {author} {\bibfnamefont {L.}~\bibnamefont {Fu}},\ }\bibfield
  {title} {\bibinfo {title} {Unconventional superconductivity and density waves
  in twisted bilayer graphene},\ }\href
  {https://doi.org/10.1103/PhysRevX.8.041041} {\bibfield  {journal} {\bibinfo
  {journal} {Phys. Rev. X}\ }\textbf {\bibinfo {volume} {8}},\ \bibinfo {pages}
  {041041} (\bibinfo {year} {2018})}\BibitemShut {NoStop}%
\bibitem [{\citenamefont {Wang}\ \emph {et~al.}()\citenamefont {Wang},
  \citenamefont {Kang},\ and\ \citenamefont {Fernandes}}]{Wang2020arxiv}%
  \BibitemOpen
  \bibfield  {author} {\bibinfo {author} {\bibfnamefont {Y.}~\bibnamefont
  {Wang}}, \bibinfo {author} {\bibfnamefont {J.}~\bibnamefont {Kang}},\ and\
  \bibinfo {author} {\bibfnamefont {R.~M.}\ \bibnamefont {Fernandes}},\
  }\bibfield  {title} {\bibinfo {title} {Topological and nematic
  superconductivity mediated by ferro-{SU(4)} fluctuations in twisted bilayer
  graphene},\ }\href@noop {} {\ }\Eprint {https://arxiv.org/abs/2009.01237}
  {2009.01237} \BibitemShut {NoStop}%
\bibitem [{\citenamefont {Lin}\ and\ \citenamefont
  {Nandkishore}(2019)}]{Lin2019highTc}%
  \BibitemOpen
  \bibfield  {author} {\bibinfo {author} {\bibfnamefont {Y.-P.}\ \bibnamefont
  {Lin}}\ and\ \bibinfo {author} {\bibfnamefont {R.~M.}\ \bibnamefont
  {Nandkishore}},\ }\bibfield  {title} {\bibinfo {title} {Chiral twist on the
  high-${T}_{c}$ phase diagram in moir\'e heterostructures},\ }\href
  {https://doi.org/10.1103/PhysRevB.100.085136} {\bibfield  {journal} {\bibinfo
   {journal} {Phys. Rev. B}\ }\textbf {\bibinfo {volume} {100}},\ \bibinfo
  {pages} {085136} (\bibinfo {year} {2019})}\BibitemShut {NoStop}%
\bibitem [{\citenamefont {Gonz\'alez}\ and\ \citenamefont
  {Stauber}(2019)}]{Gonzalez2019}%
  \BibitemOpen
  \bibfield  {author} {\bibinfo {author} {\bibfnamefont {J.}~\bibnamefont
  {Gonz\'alez}}\ and\ \bibinfo {author} {\bibfnamefont {T.}~\bibnamefont
  {Stauber}},\ }\bibfield  {title} {\bibinfo {title} {Kohn-luttinger
  superconductivity in twisted bilayer graphene},\ }\href
  {https://doi.org/10.1103/PhysRevLett.122.026801} {\bibfield  {journal}
  {\bibinfo  {journal} {Phys. Rev. Lett.}\ }\textbf {\bibinfo {volume} {122}},\
  \bibinfo {pages} {026801} (\bibinfo {year} {2019})}\BibitemShut {NoStop}%
\bibitem [{\citenamefont {Ashcroft}\ and\ \citenamefont
  {Mermin}(1976)}]{Ashcroft-book}%
  \BibitemOpen
  \bibfield  {author} {\bibinfo {author} {\bibfnamefont {N.~W.}\ \bibnamefont
  {Ashcroft}}\ and\ \bibinfo {author} {\bibfnamefont {N.~D.}\ \bibnamefont
  {Mermin}},\ }\href@noop {} {\emph {\bibinfo {title} {Solid State Physics}}}\
  (\bibinfo  {publisher} {Saunders College Publishing, Fort Worth,},\ \bibinfo
  {year} {1976})\BibitemShut {NoStop}%
\bibitem [{\citenamefont {G\'omez-Santos}\ and\ \citenamefont
  {Stauber}(2011)}]{Gomez-Santos2011PRL}%
  \BibitemOpen
  \bibfield  {author} {\bibinfo {author} {\bibfnamefont {G.}~\bibnamefont
  {G\'omez-Santos}}\ and\ \bibinfo {author} {\bibfnamefont {T.}~\bibnamefont
  {Stauber}},\ }\bibfield  {title} {\bibinfo {title} {Measurable lattice
  effects on the charge and magnetic response in graphene},\ }\href
  {https://doi.org/10.1103/PhysRevLett.106.045504} {\bibfield  {journal}
  {\bibinfo  {journal} {Phys. Rev. Lett.}\ }\textbf {\bibinfo {volume} {106}},\
  \bibinfo {pages} {045504} (\bibinfo {year} {2011})}\BibitemShut {NoStop}%
\bibitem [{\citenamefont {Raoux}\ \emph {et~al.}(2015)\citenamefont {Raoux},
  \citenamefont {Pi\'echon}, \citenamefont {Fuchs},\ and\ \citenamefont
  {Montambaux}}]{Raoux2015PRB}%
  \BibitemOpen
  \bibfield  {author} {\bibinfo {author} {\bibfnamefont {A.}~\bibnamefont
  {Raoux}}, \bibinfo {author} {\bibfnamefont {F.}~\bibnamefont {Pi\'echon}},
  \bibinfo {author} {\bibfnamefont {J.-N.}\ \bibnamefont {Fuchs}},\ and\
  \bibinfo {author} {\bibfnamefont {G.}~\bibnamefont {Montambaux}},\ }\bibfield
   {title} {\bibinfo {title} {Orbital magnetism in coupled-bands models},\
  }\href {https://doi.org/10.1103/PhysRevB.91.085120} {\bibfield  {journal}
  {\bibinfo  {journal} {Phys. Rev. B}\ }\textbf {\bibinfo {volume} {91}},\
  \bibinfo {pages} {085120} (\bibinfo {year} {2015})}\BibitemShut {NoStop}%
\bibitem [{\citenamefont {Li}\ \emph {et~al.}()\citenamefont {Li},
  \citenamefont {Jin}, \citenamefont {Yang},\ and\ \citenamefont
  {Yao}}]{Li2020tg}%
  \BibitemOpen
  \bibfield  {author} {\bibinfo {author} {\bibfnamefont {J.}~\bibnamefont
  {Li}}, \bibinfo {author} {\bibfnamefont {S.}~\bibnamefont {Jin}}, \bibinfo
  {author} {\bibfnamefont {F.}~\bibnamefont {Yang}},\ and\ \bibinfo {author}
  {\bibfnamefont {D.-X.}\ \bibnamefont {Yao}},\ }\bibfield  {title} {\bibinfo
  {title} {Electronic structure, magnetism and high-temperature
  superconductivity in the multi-layer octagraphene and octagraphite},\
  }\href@noop {} {\ }\Eprint {https://arxiv.org/abs/2008.09620v1}
  {2008.09620v1} \BibitemShut {NoStop}%
\bibitem [{\citenamefont {Louvet}\ \emph {et~al.}(2015)\citenamefont {Louvet},
  \citenamefont {Delplace}, \citenamefont {Fedorenko},\ and\ \citenamefont
  {Carpentier}}]{Louvet2015}%
  \BibitemOpen
  \bibfield  {author} {\bibinfo {author} {\bibfnamefont {T.}~\bibnamefont
  {Louvet}}, \bibinfo {author} {\bibfnamefont {P.}~\bibnamefont {Delplace}},
  \bibinfo {author} {\bibfnamefont {A.~A.}\ \bibnamefont {Fedorenko}},\ and\
  \bibinfo {author} {\bibfnamefont {D.}~\bibnamefont {Carpentier}},\ }\bibfield
   {title} {\bibinfo {title} {On the origin of minimal conductivity at a band
  crossing},\ }\href {https://doi.org/10.1103/physrevb.92.155116} {\bibfield
  {journal} {\bibinfo  {journal} {Physical Review B}\ }\textbf {\bibinfo
  {volume} {92}},\ \bibinfo {pages} {155116} (\bibinfo {year}
  {2015})}\BibitemShut {NoStop}%
\bibitem [{\citenamefont {Löwdin}(1951)}]{Lowdin1951}%
  \BibitemOpen
  \bibfield  {author} {\bibinfo {author} {\bibfnamefont {P.-O.}\ \bibnamefont
  {Löwdin}},\ }\bibfield  {title} {\bibinfo {title} {A note on the
  quantum-mechanical perturbation theory},\ }\href
  {https://doi.org/10.1063/1.1748067} {\bibfield  {journal} {\bibinfo
  {journal} {The Journal of Chemical Physics}\ }\textbf {\bibinfo {volume}
  {19}},\ \bibinfo {pages} {1396} (\bibinfo {year} {1951})}\BibitemShut
  {NoStop}%
\bibitem [{\citenamefont {Lim}\ \emph {et~al.}(2020)\citenamefont {Lim},
  \citenamefont {Fuchs}, \citenamefont {Pi\'echon},\ and\ \citenamefont
  {Montambaux}}]{Lim2020PRB}%
  \BibitemOpen
  \bibfield  {author} {\bibinfo {author} {\bibfnamefont {L.-K.}\ \bibnamefont
  {Lim}}, \bibinfo {author} {\bibfnamefont {J.-N.}\ \bibnamefont {Fuchs}},
  \bibinfo {author} {\bibfnamefont {F.}~\bibnamefont {Pi\'echon}},\ and\
  \bibinfo {author} {\bibfnamefont {G.}~\bibnamefont {Montambaux}},\ }\bibfield
   {title} {\bibinfo {title} {Dirac points emerging from flat bands in
  {L}ieb-kagome lattices},\ }\href
  {https://doi.org/10.1103/PhysRevB.101.045131} {\bibfield  {journal} {\bibinfo
   {journal} {Phys. Rev. B}\ }\textbf {\bibinfo {volume} {101}},\ \bibinfo
  {pages} {045131} (\bibinfo {year} {2020})}\BibitemShut {NoStop}%
\bibitem [{\citenamefont {Pi\'echon}\ \emph {et~al.}(2016)\citenamefont
  {Pi\'echon}, \citenamefont {Raoux}, \citenamefont {Fuchs},\ and\
  \citenamefont {Montambaux}}]{Piechon2016PRB}%
  \BibitemOpen
  \bibfield  {author} {\bibinfo {author} {\bibfnamefont {F.}~\bibnamefont
  {Pi\'echon}}, \bibinfo {author} {\bibfnamefont {A.}~\bibnamefont {Raoux}},
  \bibinfo {author} {\bibfnamefont {J.-N.}\ \bibnamefont {Fuchs}},\ and\
  \bibinfo {author} {\bibfnamefont {G.}~\bibnamefont {Montambaux}},\ }\bibfield
   {title} {\bibinfo {title} {Geometric orbital susceptibility: Quantum metric
  without {B}erry curvature},\ }\href
  {https://doi.org/10.1103/PhysRevB.94.134423} {\bibfield  {journal} {\bibinfo
  {journal} {Phys. Rev. B}\ }\textbf {\bibinfo {volume} {94}},\ \bibinfo
  {pages} {134423} (\bibinfo {year} {2016})}\BibitemShut {NoStop}%
\bibitem [{\citenamefont {Fukuyama}(1971)}]{Fukuyama1971}%
  \BibitemOpen
  \bibfield  {author} {\bibinfo {author} {\bibfnamefont {H.}~\bibnamefont
  {Fukuyama}},\ }\bibfield  {title} {\bibinfo {title} {Theory of orbital
  magnetism of {B}loch electrons: Coulomb interactions},\ }\href
  {https://doi.org/10.1143/ptp.45.704} {\bibfield  {journal} {\bibinfo
  {journal} {Progress of Theoretical Physics}\ }\textbf {\bibinfo {volume}
  {45}},\ \bibinfo {pages} {704} (\bibinfo {year} {1971})}\BibitemShut
  {NoStop}%
\bibitem [{\citenamefont {Raoux}\ \emph {et~al.}(2014)\citenamefont {Raoux},
  \citenamefont {Morigi}, \citenamefont {Fuchs}, \citenamefont
  {Pi{\'{e}}chon},\ and\ \citenamefont {Montambaux}}]{Raoux2014}%
  \BibitemOpen
  \bibfield  {author} {\bibinfo {author} {\bibfnamefont {A.}~\bibnamefont
  {Raoux}}, \bibinfo {author} {\bibfnamefont {M.}~\bibnamefont {Morigi}},
  \bibinfo {author} {\bibfnamefont {J.-N.}\ \bibnamefont {Fuchs}}, \bibinfo
  {author} {\bibfnamefont {F.}~\bibnamefont {Pi{\'{e}}chon}},\ and\ \bibinfo
  {author} {\bibfnamefont {G.}~\bibnamefont {Montambaux}},\ }\bibfield  {title}
  {\bibinfo {title} {From dia- to paramagnetic orbital susceptibility of
  massless fermions},\ }\href {https://doi.org/10.1103/physrevlett.112.026402}
  {\bibfield  {journal} {\bibinfo  {journal} {Physical Review Letters}\
  }\textbf {\bibinfo {volume} {112}},\ \bibinfo {pages} {026402} (\bibinfo
  {year} {2014})}\BibitemShut {NoStop}%
\bibitem [{\citenamefont {Landau}(1930)}]{Landau1930}%
  \BibitemOpen
  \bibfield  {author} {\bibinfo {author} {\bibfnamefont {L.}~\bibnamefont
  {Landau}},\ }\bibfield  {title} {\bibinfo {title} {Diamagnetismus der
  metalle},\ }\href {https://doi.org/10.1007/bf01397213} {\bibfield  {journal}
  {\bibinfo  {journal} {Zeitschrift f\"{u}r Physik}\ }\textbf {\bibinfo
  {volume} {64}},\ \bibinfo {pages} {629} (\bibinfo {year} {1930})}\BibitemShut
  {NoStop}%
\bibitem [{\citenamefont {Peierls}(1933)}]{Peierls1933}%
  \BibitemOpen
  \bibfield  {author} {\bibinfo {author} {\bibfnamefont {R.}~\bibnamefont
  {Peierls}},\ }\bibfield  {title} {\bibinfo {title} {Zur theorie des
  diamagnetismus von leitungselektronen},\ }\href
  {https://doi.org/10.1007/bf01342591} {\bibfield  {journal} {\bibinfo
  {journal} {Zeitschrift fur Physik}\ }\textbf {\bibinfo {volume} {80}},\
  \bibinfo {pages} {763} (\bibinfo {year} {1933})}\BibitemShut {NoStop}%
\bibitem [{\citenamefont {Landau}\ and\ \citenamefont
  {Lifshitz}(1981)}]{Landau-course9}%
  \BibitemOpen
  \bibfield  {author} {\bibinfo {author} {\bibfnamefont {L.~D.}\ \bibnamefont
  {Landau}}\ and\ \bibinfo {author} {\bibfnamefont {E.~M.}\ \bibnamefont
  {Lifshitz}},\ }\href@noop {} {\emph {\bibinfo {title} {Course of Theoretical
  Physics}}},\ Vol.\ \bibinfo {volume} {9, Pt. 2, Sec.57.}\ (\bibinfo
  {publisher} {Pergamon, New York,},\ \bibinfo {year} {1981})\BibitemShut
  {NoStop}%
\bibitem [{Sup()}]{Supplement}%
  \BibitemOpen
  \href@noop {} {\bibinfo {title} {See supplemental material at
  [\href{http://link.aps.org/supplemental/10.1103/PhysRevB.103.195104}{http://link.aps.org/supplemental/10.1103/PhysRevB.103.195104}]
  for more detailed evolution of orbital susceptibility as a function of
  chemical potential with changing hopping ratio $\alpha$.}}\BibitemShut
  {Stop}%
\bibitem [{\citenamefont {Pi{\'{e}}chon}\ \emph {et~al.}(2015)\citenamefont
  {Pi{\'{e}}chon}, \citenamefont {Fuchs}, \citenamefont {Raoux},\ and\
  \citenamefont {Montambaux}}]{Piechon2015JPCM}%
  \BibitemOpen
  \bibfield  {author} {\bibinfo {author} {\bibfnamefont {F.}~\bibnamefont
  {Pi{\'{e}}chon}}, \bibinfo {author} {\bibfnamefont {J.-N.}\ \bibnamefont
  {Fuchs}}, \bibinfo {author} {\bibfnamefont {A.}~\bibnamefont {Raoux}},\ and\
  \bibinfo {author} {\bibfnamefont {G.}~\bibnamefont {Montambaux}},\ }\bibfield
   {title} {\bibinfo {title} {Tunable orbital susceptibility in $\alpha-t_3$
  tight-binding models},\ }\href
  {https://doi.org/10.1088/1742-6596/603/1/012001} {\bibfield  {journal}
  {\bibinfo  {journal} {Journal of Physics: Conference Series}\ }\textbf
  {\bibinfo {volume} {603}},\ \bibinfo {pages} {012001} (\bibinfo {year}
  {2015})}\BibitemShut {NoStop}%
\bibitem [{\citenamefont {McClure}(1956)}]{McClure1956}%
  \BibitemOpen
  \bibfield  {author} {\bibinfo {author} {\bibfnamefont {J.~W.}\ \bibnamefont
  {McClure}},\ }\bibfield  {title} {\bibinfo {title} {Diamagnetism of
  graphite},\ }\href {https://doi.org/10.1103/physrev.104.666} {\bibfield
  {journal} {\bibinfo  {journal} {Physical Review}\ }\textbf {\bibinfo {volume}
  {104}},\ \bibinfo {pages} {666} (\bibinfo {year} {1956})}\BibitemShut
  {NoStop}%
\bibitem [{\citenamefont {Rosenzweig}\ \emph {et~al.}()\citenamefont
  {Rosenzweig}, \citenamefont {Karakachian}, \citenamefont {Marchenko},
  \citenamefont {Küster},\ and\ \citenamefont {Starke}}]{Rosenzweig2020}%
  \BibitemOpen
  \bibfield  {author} {\bibinfo {author} {\bibfnamefont {P.}~\bibnamefont
  {Rosenzweig}}, \bibinfo {author} {\bibfnamefont {H.}~\bibnamefont
  {Karakachian}}, \bibinfo {author} {\bibfnamefont {D.}~\bibnamefont
  {Marchenko}}, \bibinfo {author} {\bibfnamefont {K.}~\bibnamefont {Küster}},\
  and\ \bibinfo {author} {\bibfnamefont {U.}~\bibnamefont {Starke}},\
  }\bibfield  {title} {\bibinfo {title} {Overdoping graphene beyond the van
  {H}ove singularity},\ }\href@noop {} {\ }\Eprint
  {https://arxiv.org/abs/2009.04876} {2009.04876} \BibitemShut {NoStop}%
\bibitem [{\citenamefont {Choudhry}\ \emph {et~al.}(2019)\citenamefont
  {Choudhry}, \citenamefont {Yue},\ and\ \citenamefont
  {Liao}}]{Choudhry2019PRB}%
  \BibitemOpen
  \bibfield  {author} {\bibinfo {author} {\bibfnamefont {U.}~\bibnamefont
  {Choudhry}}, \bibinfo {author} {\bibfnamefont {S.}~\bibnamefont {Yue}},\ and\
  \bibinfo {author} {\bibfnamefont {B.}~\bibnamefont {Liao}},\ }\bibfield
  {title} {\bibinfo {title} {Origins of significant reduction of lattice
  thermal conductivity in graphene allotropes},\ }\href
  {https://doi.org/10.1103/PhysRevB.100.165401} {\bibfield  {journal} {\bibinfo
   {journal} {Phys. Rev. B}\ }\textbf {\bibinfo {volume} {100}},\ \bibinfo
  {pages} {165401} (\bibinfo {year} {2019})}\BibitemShut {NoStop}%
\bibitem [{\citenamefont {Guerci}\ \emph {et~al.}(2021)\citenamefont {Guerci},
  \citenamefont {Simon},\ and\ \citenamefont {Mora}}]{Guerci2021}%
  \BibitemOpen
  \bibfield  {author} {\bibinfo {author} {\bibfnamefont {D.}~\bibnamefont
  {Guerci}}, \bibinfo {author} {\bibfnamefont {P.}~\bibnamefont {Simon}},\ and\
  \bibinfo {author} {\bibfnamefont {C.}~\bibnamefont {Mora}},\ }\bibfield
  {title} {\bibinfo {title} {Moir\'{e} lattice effects on the orbital magnetic
  response of twisted bilayer graphene and condon instability},\ }\href@noop {}
  {\  (\bibinfo {year} {2021})},\ \Eprint {https://arxiv.org/abs/2103.13459}
  {arXiv:2103.13459 [cond-mat.mes-hall]} \BibitemShut {NoStop}%
\bibitem [{\citenamefont {Shen}\ \emph {et~al.}(2010)\citenamefont {Shen},
  \citenamefont {Shao}, \citenamefont {Wang},\ and\ \citenamefont
  {Xing}}]{Shen2010PRB}%
  \BibitemOpen
  \bibfield  {author} {\bibinfo {author} {\bibfnamefont {R.}~\bibnamefont
  {Shen}}, \bibinfo {author} {\bibfnamefont {L.~B.}\ \bibnamefont {Shao}},
  \bibinfo {author} {\bibfnamefont {B.}~\bibnamefont {Wang}},\ and\ \bibinfo
  {author} {\bibfnamefont {D.~Y.}\ \bibnamefont {Xing}},\ }\bibfield  {title}
  {\bibinfo {title} {Single {D}irac cone with a flat band touching on
  line-centered-square optical lattices},\ }\href
  {https://doi.org/10.1103/PhysRevB.81.041410} {\bibfield  {journal} {\bibinfo
  {journal} {Phys. Rev. B}\ }\textbf {\bibinfo {volume} {81}},\ \bibinfo
  {pages} {041410(R)} (\bibinfo {year} {2010})}\BibitemShut {NoStop}%
\bibitem [{\citenamefont {Green}\ \emph {et~al.}(2010)\citenamefont {Green},
  \citenamefont {Santos},\ and\ \citenamefont {Chamon}}]{Green2010PRB}%
  \BibitemOpen
  \bibfield  {author} {\bibinfo {author} {\bibfnamefont {D.}~\bibnamefont
  {Green}}, \bibinfo {author} {\bibfnamefont {L.}~\bibnamefont {Santos}},\ and\
  \bibinfo {author} {\bibfnamefont {C.}~\bibnamefont {Chamon}},\ }\bibfield
  {title} {\bibinfo {title} {Isolated flat bands and spin-1 conical bands in
  two-dimensional lattices},\ }\href
  {https://doi.org/10.1103/PhysRevB.82.075104} {\bibfield  {journal} {\bibinfo
  {journal} {Phys. Rev. B}\ }\textbf {\bibinfo {volume} {82}},\ \bibinfo
  {pages} {075104} (\bibinfo {year} {2010})}\BibitemShut {NoStop}%
\bibitem [{\citenamefont {Gorbar}\ \emph {et~al.}(2019)\citenamefont {Gorbar},
  \citenamefont {Gusynin},\ and\ \citenamefont {Oriekhov}}]{Gorbar2019PRB}%
  \BibitemOpen
  \bibfield  {author} {\bibinfo {author} {\bibfnamefont {E.~V.}\ \bibnamefont
  {Gorbar}}, \bibinfo {author} {\bibfnamefont {V.~P.}\ \bibnamefont
  {Gusynin}},\ and\ \bibinfo {author} {\bibfnamefont {D.~O.}\ \bibnamefont
  {Oriekhov}},\ }\bibfield  {title} {\bibinfo {title} {Electron states for
  gapped pseudospin-1 fermions in the field of a charged impurity},\ }\href
  {https://doi.org/10.1103/physrevb.99.155124} {\bibfield  {journal} {\bibinfo
  {journal} {Physical Review B}\ }\textbf {\bibinfo {volume} {99}},\ \bibinfo
  {pages} {155124} (\bibinfo {year} {2019})}\BibitemShut {NoStop}%
\bibitem [{\citenamefont {Oriekhov}\ \emph {et~al.}(2018)\citenamefont
  {Oriekhov}, \citenamefont {Gorbar},\ and\ \citenamefont
  {Gusynin}}]{Oriekhov2018LTP}%
  \BibitemOpen
  \bibfield  {author} {\bibinfo {author} {\bibfnamefont {D.~O.}\ \bibnamefont
  {Oriekhov}}, \bibinfo {author} {\bibfnamefont {E.~V.}\ \bibnamefont
  {Gorbar}},\ and\ \bibinfo {author} {\bibfnamefont {V.~P.}\ \bibnamefont
  {Gusynin}},\ }\bibfield  {title} {\bibinfo {title} {Electronic states of
  pseudospin-1 fermions in dice lattice ribbon},\ }\href
  {https://doi.org/10.1063/1.5078627} {\bibfield  {journal} {\bibinfo
  {journal} {Low Temperature Physics}\ }\textbf {\bibinfo {volume} {44}},\
  \bibinfo {pages} {1313} (\bibinfo {year} {2018})}\BibitemShut {NoStop}%
\end{thebibliography}%

\end{document}